\documentclass[hyper]{JHEP3}

\pdfoutput=1

\usepackage{enumerate}
\usepackage{graphicx}
\usepackage{epsfig}
\usepackage{amsmath,amssymb,amsfonts,mathrsfs}
\usepackage{slashed}

\newcommand{\bea}{\begin{eqnarray}}
\newcommand{\eea}{\end{eqnarray}}
\newcommand{\bean}{\begin{eqnarray*}}
\newcommand{\eean}{\end{eqnarray*}}

\def\beq{\begin{equation}}
\def\be{\begin{equation}}
\def\eeq{\end{equation}}
\def\ee{\end{equation}}
\def\half{\frac{1}{2}}

\title{Discovery potential of top-partners in a realistic composite Higgs model with early LHC data}

\author{G\"unther Dissertori\\
  Institute for Particle Physics, ETH Zurich,\\
  8093 Zurich, Switzerland\\
  E-mail: \email{dissertori@phys.ethz.ch}}
\author{Elisabetta Furlan\\
  Institute for Theoretical Physics, ETH Zurich,\\
  8093 Zurich, Switzerland\\
  E-mail: \email{efurlan@phys.ethz.ch}}
\author{Filip Moortgat\\
  Institute for Particle Physics, ETH Zurich,\\
  8093 Zurich, Switzerland\\
  E-mail: \email{filip.moortgat@cern.ch}}
\author{Pascal Nef\\
  Institute for Particle Physics, ETH Zurich,\\
  8093 Zurich, Switzerland\\
  E-mail: \email{pascal.nef@cern.ch}}

\abstract{Composite Higgs models provide a natural, non-supersymmetric solution to the
hierarchy problem. In these models, one or more sets of heavy top-partners are typically 
introduced. Some of these new quarks can be relatively light, with a mass of a few hundred 
GeV, and could be observed with the early LHC collision data expected to be collected during 
2010. We analyse in detail the collider signatures that these new quarks can produce. We 
show that final states with two (same-sign) or three leptons are the most promising discovery 
channels. They can yield a $5 \sigma$ excess over the Standard Model expectation already with the 2010 
LHC collision data. Exotic quarks of charge 5/3 are a distinctive feature of this model.
We present a new method to reconstruct their masses from their leptonic decay without
relying on jets in the final state.
}


\preprint{  }

\begin{document}

\section{Introduction}
\label{sec:introduction}

The Standard Model (SM) of particle physics has so far been 
experimentally confirmed in many of its aspects. Yet, a fundamental 
piece is still missing; namely, the understanding of the 
mechanism responsible for the breaking of the $SU(2)_L \times U(1)_Y $ electroweak 
symmetry. The `minimal' description provided in the SM 
consists in the introduction of a complex scalar doublet $\varphi$. 
Electroweak symmetry breaking (EWSB) is achieved assuming that 
this field acquires a non-zero vacuum expectation value. 
After EWSB, only one physical degree of freedom survives: the 
Higgs boson. Experimental results point towards a relatively light 
particle. If the Higgs boson exists, it should be within reach of the 
LHC.

A light fundamental scalar is not natural, though. Radiative corrections are 
expected to drive its mass close to the Planck scale (or to the scale of 
onset of some new physics). 
An elegant way to prevent this is through symmetries. The most famous 
example is supersymmetry, that exploits the cancellation between 
the contributions given by fermions and by bosons to the Higgs self-energy. 
This is not the only solution. 
Composite Higgs models~\cite{Kaplan:1983fs, Kaplan:1983sm} 
provide an alternative mechanism to explain the lightness of the Higgs boson. 
In these models the Higgs boson arises as a composite state 
of some new, strongly interacting sector. The new sector 
possesses a global symmetry that is spontaneously broken
at some scale $f$. 
This symmetry breaking gives (at least) four Goldstone bosons 
that can be arranged into a complex $SU(2)_L$ doublet, which 
we identify with the Higgs doublet. 
Upon gauging the electroweak symmetry group, the Higgs boson acquires a potential, and hence a mass.
Since we are interested in the low-energy regime of this 
strongly coupled theory, we can adopt an effective Lagrangian 
approach~\cite{Giudice:2007fh}. 
We will consider the minimal symmetry breaking pattern
$SO(5)/SO(4)$~\cite{Agashe:2004rs}, 
that also preserves custodial symmetry.

The Higgs boson is not necessarily the only composite state of the 
new sector to be relatively light.
In particular, the mixing of the top with composite quarks 
can explain the large top mass. These composite quarks 
can give significant contributions to the electroweak precision observables, 
thus modifying the region of parameter space that is allowed for these models
~\cite{Carena:2006bn,Barbieri:2007bh,Carena:2007ua,Lodone:2008yy,Gillioz:2008hs,Anastasiou:2009rv}. 
For this study, we focus on a non-minimal realization of this model, 
where two multiplets of top-partners in the fundamental representation of $SO(5)$
are introduced~\cite{Anastasiou:2009rv}.

The LHC is expected to run throughout this year 
at a center of mass energy \mbox{$\sqrt{s}=7$ TeV}, opening 
an unprecedented window for searches of new phenomena in particle physics.
The first glimpse of new physics could well be due to new heavy quarks, which are 
a rather common feature of Beyond the Standard Model (BSM) scenarios. 
The discovery potential of such heavy quarks has been studied in the context of
little and littlest Higgs models~\cite{Azuelos:2004dm, Karafasoulis:962029, PhysRevD.75.055006},
warped extra dimensions~\cite{Carena:2007tn, Dennis:2007tv}, 
fourth generation quarks~\cite{Holdom:2007ap,Holdom:2007nw, CMS-PAS-EXO-08-013, CMS-PAS-EXO-09-012} 
and generic vector-like quarks in isospin singlets or doublets and with different hypercharge~\cite{AguilarSaavedra:2009es}. 
If new quarks are observed, we will need a way to understand 
which model they point at. 
For this reason we focus on collider signatures that can be considered distinctive of the 
composite Higgs model under study. In particular, we look for configurations in which 
either two charge 5/3 quarks or a full, almost degenerate $\mathbf{4}$ of $SO(4)$ 
lie within the reach of the 2010 runs at the LHC. 
For these distinctive signatures, we discuss the phenomenology and 
study the discovery potential
on the basis of $200 \ \mathrm{pb}^{-1}$ of 
collision data at $\sqrt{s}=7 \ \mathrm{TeV}$. We study the event yield with
respect to the SM expectation in various multi-lepton channels for 
different points in the parameter space.
We outline a new method to reconstruct the mass
of a charge 5/3 top-partner exploiting its leptonic decay channel.
We show that with only about 50 signal events in the same-sign di-lepton final state,
this method can be used to judge if the signal is mainly due to one charge 5/3 quark
or rather produced by the contributions from multiple top-partners.  

The paper is organized as follows. In section~\ref{sec:model} we review 
the composite Higgs model of ref.~\cite{Anastasiou:2009rv}. 
In section~\ref{sec:phenomenology} we discuss the general features of the
phenomenology of the two distinctive signatures of the model.
In section~\ref{sec:implementation_and_grouping} 
we describe how the model was implemented in an event-generator to allow for consistent 
event generation within a specific point in parameter space.
The generation of signal and background 
samples and the fast detector simulation are discussed in 
section~\ref{sec:Simulation_and_Reconstruction}.
Section~\ref{sec:discovery} is dedicated to the description of the discovery potential 
in multi-lepton final states. 
We focus on two particularly interesting points and discuss 
their phenomenology and the discovery potential by means of a robust cut-based analysis.
In section~\ref{sec:mass_reco}, we present a new method 
to reconstruct the mass of a charge 5/3 top-partner via its leptonic decay.


\section{The model}
\label{sec:model}
\subsection{The Higgs sector}
	We consider a strongly interacting sector that can be described at 
	low energy by a non-linear sigma model. The cutoff of this model
	is 
	$\Lambda_{UV}  = 4 \pi f/ \sqrt{N_G}$, 
	where $N_G$ is the number of Goldstone bosons and $f$ is the scale 
	at which the $SO(5) \rightarrow SO(4)$ breaking occurs. 
	This scale is assumed to be larger than the EWSB scale $v = 174$~GeV.
	Too large values of $f$ would introduce a substantial fine-tuning of the 
	model~\cite{Barbieri:2007bh}; on the other hand, if the scale of new physics is too low, 
	large contributions to electroweak parameters and flavour physics are 
	introduced. For these reasons we set $f=500$~GeV, which corresponds to a 
	$\sim 10\%$ fine-tuning~\cite{Barbieri:2007bh}.
	
	The $SO(5) \rightarrow SO(4)$ breaking 
	is realized  through a scalar $\phi$ subject to the constraint 
	$$
	\phi^2 = f^2 \;.
	$$
	In the non-linear representation 
	\beq
		\phi  = \phi_0 e^{
										-i T^{\hat{a}} h_{\hat{a}} \sqrt{2}/f}\;,
	\eeq
	where 
	$\phi_0 = (0, 0, 0, 0, f)$
	is the vacuum state that preserves $SO(4)$,  
	$T^{\hat{a}}$ are the four broken generators 
	and $h_{\hat{a}}$ the corresponding Goldstone bosons. 
	Expanding the exponential, we get
	\beq
		\phi = f \left(
						\frac{	h_{ \hat{a} }	}
								{	h	}
							\sin \frac{ h }{ f }
							,
							\cos \frac{ h }{ f }
						\right)
				\equiv
					\left(
							\vec{\phi},
							\phi_5
					\right)
					\;,
	\eeq
	where
	$ \vec{\phi} \equiv (\tilde{\varphi}, \varphi)$ 
	transforms under the fundamental representation of 
	$ SO(4)\equiv SU(2)_L \times SU(2)_R $  and 
	$ h 	= 	\sqrt{h_{ \hat{a} }^2} $. 
	We denote by 
	$ \varphi $ and $\tilde{\varphi}$ 
	the SM Higgs doublets with hypercharge +1/2 and -1/2. 
	Finally, we gauge $SU(2)_L$ and the $T^3_R$ generator of $SU(2)_R$. 
	This explicitly
	breaks the $SO(5)$ symmetry and induces a potential for the Higgs boson, that becomes
	a pseudo-Goldstone boson. Since the potential is generated at loop level, the mass of the Higgs boson
	is expected to be light. Throughout this
	study we set $m_h = 120$ GeV.

	The usual relation for the mass of the $W$ boson,
	\beq
		m_W^2 = \frac{g^2 v^2}{2} 
		\quad, \quad
		v^2 = \half 
					\langle \vec{\phi}^2\rangle
					\;
	\eeq
	holds provided that we set
	\beq
		\sin \left(
						\frac{
								\sqrt{2}
								\langle \varphi \rangle
								}
								{	h	}
				\right)
		=
		\frac{ v \sqrt{2} }{f}
		\equiv
		s_{\alpha}
		\;.
	\label{eq:s_a}
	\eeq
	For $s_{\alpha} = 0$, $\phi = \phi_0$, electroweak symmetry 
	remains unbroken and the gauge bosons are massless, while
	$s_{\alpha} = 1$ corresponds to maximal EWSB. 
	
	Higgs compositeness, together with the requirement for canonical 
	normalization of the kinetic term, leads to a rescaling of the 
	physical Higgs field by a factor
	$
		c_{\alpha} = \sqrt{ 1- 2 v^2/f^2}
	$. 
	This implies an analogous reduction of the couplings between the Higgs
	and the gauge bosons and gives in turn some dependence of the 
	electroweak precision test (EWPT) observables 
	on the UV cutoff of the model. In fact, in the SM the Higgs boson regulates the logarithmic 
	divergencies of the gauge bosons self-energies. In the heavy Higgs approximation,
	the Peskin-Takeuchi parameters $S$ and $T$~\cite{Altarelli:1990zd} read
	\beq
		S,T = a_{S,T} \log{m_h} + b_{S,T}
		\;,
	\eeq
	where $m_h$ is the mass of the Higgs boson and $a_{S,T}$, $b_{S,T}$ are constants. 
	The reduction of the Higgs boson couplings to the gauge boson spoils the 
	cancellation of the logarithmic dependence on the UV cutoff, so that now
	\beq
		S,T = a_{S,T} (
							c_{\alpha}^2 \log{m_h} 
							+ s_{\alpha}^2 \log \Lambda_{UV}
							)+ b_{S,T}
		\;.
	\eeq
	This can be taken into account when one computes EWPT observables 
	by replacing the Higgs mass with an effective mass~\cite{Barbieri:2007bh}
	\beq
		m_{\mathrm{EWPT, eff}}
		=
		m_h
		\left(
				\Lambda_{UV}/m_h
		\right)^{s_{\alpha}^2}
		\;.
	\eeq
	As a consequence, we obtain an extra positive contribution to $S$ and a negative 
	contribution to $T$,
	\beq
		\Delta S = \frac{ 1 }{ 12 \pi }
						  \log \left(
						  				\frac{ m_{\mathrm{EWPT, eff}}^2 }{ m_{h,\mathrm{ref}}^2 }
						  			\right)
		  \quad , \quad
		\Delta T = - \frac{ 3 }{ 16 \pi c_W^2 }
						  \log \left(
						  				\frac{ m_{\mathrm{EWPT, eff}}^2 }{ m_{h,\mathrm{ref}}^2 }
						  			\right)
		\;,
	\eeq
	where $c_W$ is the cosine of the Weinberg angle and $m_{h,\mathrm{ref}}$ is the 
	Higgs mass used in the electroweak fit. 
	
	On top of this, one can expect the strongly coupled dynamics itself to 
	affect EWPT observables through some higher-dimensional operator. This
	model includes custodial symmetry to protect the $T$ parameter. A reasonable
	estimate of the contribution to $S$ is~\cite{Barbieri:2007bh}
	\beq
		\Delta S_{\Lambda} 
		\sim
		\frac{ 4 s_W^2 }{ \alpha_{em} }
		\frac{ g^2 v^2 }{ \Lambda^2 }
		\approx 
		0.16 \left(
						\frac{ 3\mbox{ TeV} }{ \Lambda }
				\right)^2
			\;.
	\eeq
	
	Combining the effect from Higgs compositeness and higher-order operators, 
	one typically obtains too large contributions to the $S$ and $T$ parameters 
	and the model is not compatible with current EWPT constraints
	\cite{Barbieri:2007bh, Gillioz:2008hs, Anastasiou:2009rv}. 
	Yet, one can 
	expect other composite states to be as well below the cutoff of the effective 
	theory. Here we will consider the case of fermionic resonances and analyze 
	how they can improve the agreement of the model with observations.

\subsection{The fermionic sector}
\label{sec:fermionic_sector}
	We consider 
	vector-like resonances of composite fermions 
	transforming in the fundamental representation of $SO(5)$. 
	We denote them by $\Psi^i$, with the index $i$ running over 
	the multiplets included below the cutoff. The corresponding mass Lagrangian is~\cite{Gillioz:2008hs, Anastasiou:2009rv}
	\beq
		-\mathcal{L}_\mathrm{SO(5)}
		= 
		M_i \bar{\Psi}^i \Psi^i
		+
		\frac{  y_{ij}  }{ f }
				(
				\bar{\Psi}^i \phi
				) 
				(
				\phi^{\dagger} \Psi^j)
		\;,
		\label{M_Psi}
	\eeq
	where $y_{ij}$ is a Hermitian matrix.
	Under the electroweak gauge group, $\Psi$ decomposes as
	$
		\Psi = (Q, X, T)
	$,
	where $Q$ and $X$ are $SU(2)_L$ doublets with hypercharge 
	+1/6 and +7/6 respectively, and $T$ is a $SU(2)_L$ singlet 
	with hypercharge 2/3.  The $X$ doublet
	introduces another quark of electromagnetic charge $2/3$, which can 
	mix with the top, and a quark with charge $5/3$. Such quarks are one 
	of the distinguishing features of the model. 
	The SM quarks $q_L$ and $t_R$ have the 
	same quantum numbers as $Q$ and $T$, respectively.
	The most generic interaction between the top sector and the new 
	quarks is therefore of the form
	\beq
		-\mathcal{L}_\mathrm{int}
		=
		\Delta_L^{i} \bar{q}_L Q^i_R 
		+ 
		\Delta_R^{i} \bar{T}^i_L t_R 
		+ 
		\mathrm{h.c.} \;.
		\label{mixing}
	\eeq
	Combining eqs.~(\ref{M_Psi}) and~(\ref{mixing}) we obtain the mass matrices
	\beq
		- \mathcal{L}^{2/3}
		=
			\begin{pmatrix}
					\overline{t}_L \\ \overline{Q^u}_L \\ \overline{X^u}_L \\ \overline{T}_L 
			\end{pmatrix}^T
		\begin{pmatrix}
					0	&	\Delta_L^\mathrm{T}	&	0	&	0 \\
					0	&	M+\frac{s_\alpha^2}{2}f y	&	\frac{s_\alpha^2}{2}f y	&	c_\alpha v y \\
					0 & \frac{s_\alpha^2}{2}f y  & M+\frac{s_\alpha^2}{2}f y   & c_\alpha v y \\ 
					\Delta_R  & c_\alpha v y  & c_\alpha v y  & M + c_\alpha^2f y 
		\end{pmatrix}
		\begin{pmatrix}
					t_R \\ Q^u_R \\ X^u_R \\ T_R
		\end{pmatrix}
		+
		\mathrm{h.c.}
		\label{mtop}
	\eeq
	for the quarks of charge 2/3 and
	\beq
		- \mathcal{L}^{-1/3}
		=
			\begin{pmatrix}
					\overline{b}_L \\ \overline{Q^d}_L
			\end{pmatrix}^T
		\begin{pmatrix}
					-\lambda_b v	&	\Delta_L^\mathrm{T} \\
					0	&	M \\
		\end{pmatrix}
		\begin{pmatrix}
					b_R \\ Q^d_R
		\end{pmatrix}
		+
		\mathrm{h.c.}
		\label{mbott}
	\eeq
	for the quarks of charge -1/3. The indices $u$ and $d$ denote 
	respectively the charge 2/3 and -1/3 components of the doublet 
	indicated. In the case of more fermionic resonances, the mass matrices 
	are to be understood as in block form.
	Note that in eq.~(\ref{mbott}) we introduced an explicit 
	$SO(5)$ breaking term
	\beq
		\mathcal{L}^b
		=
		\lambda_b \bar{q}_L \varphi b_R
	\label{eq:m_b}
	\eeq
	to give a mass to the bottom quark. 
	We could also generate a mass for the bottom quark in an $SO(5)$ preserving 
	fashion. For example, we could couple the bottom quark to some new multiplets 
	of $SO(5)$, as we did for the top quark. 
	This would come at the expense of introducing extra particles. 
	Since
	the mass of the bottom quark is small, we do not expect large effects from 
	bottom compositeness. We opt therefore for 
	a minimal description, in which the bottom mass is generated with the current particle 
	content of the model.
	
	The couplings of the fermions to the Higgs boson are obtained 
	expanding the second term in eq.~(\ref{M_Psi}) around the vev of 
	$\phi$. 
	For example, the couplings of the charge 2/3 quarks to the Higgs boson 
	are given by 
	\beq
		-\mathcal{L}^{h,t}
		=
		y h
			\begin{pmatrix}
					\overline{t}_L \\ \overline{Q^u}_L \\ \overline{X^u}_L \\ \overline{T}_L 
			\end{pmatrix}^T
		\begin{pmatrix}
					0	&	0	&	0	&	0 \\
					0	&	s_{\alpha} c_{\alpha}	&	s_{\alpha} c_{\alpha}	&	\frac{1-2 s_{\alpha}^2}{ \sqrt{2} } \\
					0	& s_{\alpha} c_{\alpha}		&	s_{\alpha} c_{\alpha}	&	\frac{1-2 s_{\alpha}^2}{ \sqrt{2} } \\
					0  &	\frac{1-2 s_{\alpha}^2}{ \sqrt{2} }	&	\frac{1-2 s_{\alpha}^2}{ \sqrt{2} }	&	-2 s_{\alpha} c_{\alpha}
		\end{pmatrix}
		\begin{pmatrix}
					t_R \\ Q^u_R \\ X^u_R \\ T_R
		\end{pmatrix}
		\;.
	\eeq
	Here we already included the suppression factor $c_{\alpha}$ in the Higgs couplings.

	As it was emphasized in refs.~\cite{Barbieri:2007bh, Lodone:2008yy, 
	Gillioz:2008hs, Anastasiou:2009rv,Contino:2006qr}, 
	composite Higgs models with only one set of fermionic 
	resonances below the cutoff of the effective theory are very constrained 
	from EWPT. The charge 5/3 quark is the lightest new particle predicted, 
	with a mass $m_{5/3} \lesssim 500$~GeV. Above it and rather close in mass 
	($\Delta m \lesssim 100$~GeV) is 
	a charge 2/3 quark. The other quarks are typically much heavier. 
	The most relevant collider signatures therefore come from the production
	and decay of the charge 5/3 quark. 
	These signatures have been studied in detail in
	\cite{CMS-PAS-EXO-08-008,Contino_08}.
	
	The scenario dramatically changes if we include a second set of composite fermions 
	below the cutoff~\cite{ Anastasiou:2009rv,Panico:2008bx}. Constraints from EWPT become  
	less stringent, and many different 
	mass patterns are allowed in the region accessible with first LHC data. 
	In the next sections we discuss the collider signatures and discovery potential 
	of this model.


\subsection{Parameter-space scan}
\label{sec:scans}
	
	We scan over the parameter space of the model in order to find 
	regions compatible with EWPT. From eqs.~(\ref{M_Psi})-(\ref{eq:m_b}) we see that 
	there are 6 variables parametrizing the fermionic sector with one multiplet below the 
	cutoff, and 11 
	for the case of two multiplets. In both cases, $s_{\alpha}$ is fixed through eq.~(\ref{eq:s_a}), 
	as we have $v=174$ GeV and $f=500$ GeV. We fix other two parameters in such a way to 
	obtain the measured top and bottom masses~\cite{Collaboration:2008ub, PDG} 
	$$
	 m_t = 172.4 {\rm \; GeV \qquad and \qquad} m_b = 4.2 {\rm \; GeV \;.}
	$$
	This is more easily done if we factor out of the mass Lagrangian~(\ref{mtop}) (or eq.~(\ref{mbott})
	for the bottom quark) 
	one of the parameters, say $M_1$ ($\lambda_b$). Then we diagonalize the remaining part of the 
	mass matrix and fix $M_1$ ($\lambda_b$) so that the mass of the lightest quark is $m_t$ ($m_b$).
	We are left with eight free parameters in the case of the two-multiplet model.
	We require that the resulting quarks contribute to EWPT observables 
	in such a way to make the model compatible with observations. We use the same fit 
	as in~\cite{Anastasiou:2009rv} to assess the 
	agreement between a point in parameter space and experimental constraints. 
	Furthermore, we exploit the value of $\chi^2$ that parametrizes this comparison in order to 
	drive our Vegas-based analysis~\cite{Hahn:2004fe}. The procedure is the following. 
	We use Vegas to randomly sample on 
	the eight-dimensional parameter space. For each point sampled, the value returned to Vegas as an 
	`integrand' is $1/ \chi^2$. By construction, Vegas will focus its sampling on the points that 
	lead to a higher value of the integrand $1/ \chi^2$, i.e., to a better agreement with EWPT. 
	We retain points that are compatible with EWPT at 99\% C.L. .
	
	We further refine our search asking for 
	signals which are characteristic of the two-multiplet model. 
	As we said, in the case of only one multiplet below the cutoff, the mass spectrum of the 
	new resonances is typically rather spread out. The charge 5/3 quark has a mass of some few 
	hundred GeV, while the charge -1/3 quark is very close to the cutoff. A signature of a 
	two-multiplet model would then be a charge 5/3 quark in a $\mathbf{4}$ of $SO(4)$, i.e.\
	very close in mass to two charge 2/3 and one 
	charge -1/3 quarks. We require the mass difference among these particles to be $\lesssim$ 60 GeV, 
	so that decays through off-shell gauge bosons are strongly suppressed. Another typical signature of 
	the model is the presence of both the charge 5/3 quarks. We take these two signatures as 
	neat indications of this particular model and focus on their discovery potential with early 
	LHC data.
	With 200~pb$^{-1}$ of collision data at 7 TeV, 
	a significant number of quarks with masses below $\sim 500$~GeV should be produced\footnote{
	For reasons that we will explain later, we focus on decay channels which 
	produce at least two charged leptons in the final state. We estimate a leading order 
	cross section of $207.8\pm 0.5$~fb
	for pair production of a quark with a mass of 500~GeV 
	at $\sqrt{s}=7$~TeV (the stated uncertainty is due to statistics only). 
	Taking into account the branching ratio for the $W$ and $Z$ bosons to decay leptonically, 
	we cannot expect to observe more than a handful of events in the considered channels
	for an integrated luminosity of $200 \ \mathrm{pb}^{-1}$.}.
	We will set this value 
	as an upper bound in our search for the two distinctive patterns that we just discussed. 
	Direct searches have set lower bounds for the mass of new quarks. 
	We use the most recent results from Tevatron 
	on the exclusion of a charge 5/3 top-partner~\cite{Aaltonen:2009nr}.
	For this quark, the only decay channel is $tW^{+}$, as in the reference.
	We do not use instead the most stringent exclusion bounds on 
	the charge -1/3 and 2/3 quarks,~\cite{Aaltonen:2009nr,Aaltonen:2007je}
	and~\cite{CDF:2008nf}, as they assume the new quarks 
	to decay entirely through either $W$ or $Z$. This is not the case 
	in our model. Therefore, as lower mass bounds we set
	$$
	m_{5/3} > 365 {\rm \; GeV \qquad, \qquad } m_{2/3},\, m_{-1/3} > 260 {\rm \; GeV \;.}
	$$

\section{Phenomenology of the two-multiplet composite Higgs model}
\label{sec:phenomenology}

	In this section, we outline some of the basic features of the phenomenology that we expect from the two-multiplet
	model. This phenomenology is largely determined by the mass hierarchy of the 10 new
	quarks. The mass eigenstates 
	(ordered according to increasing mass) of the new top-like quarks 
	will be named $t_1$, $t_2$, $t_3$, $t_4$, $t_5$ and $t_6$, whereas the charge 5/3 quarks 
	and the bottom-like quarks will be denoted as $x_1$, $x_2$ and $b_1$, $b_2$, respectively.
	
	\DOUBLEFIGURE{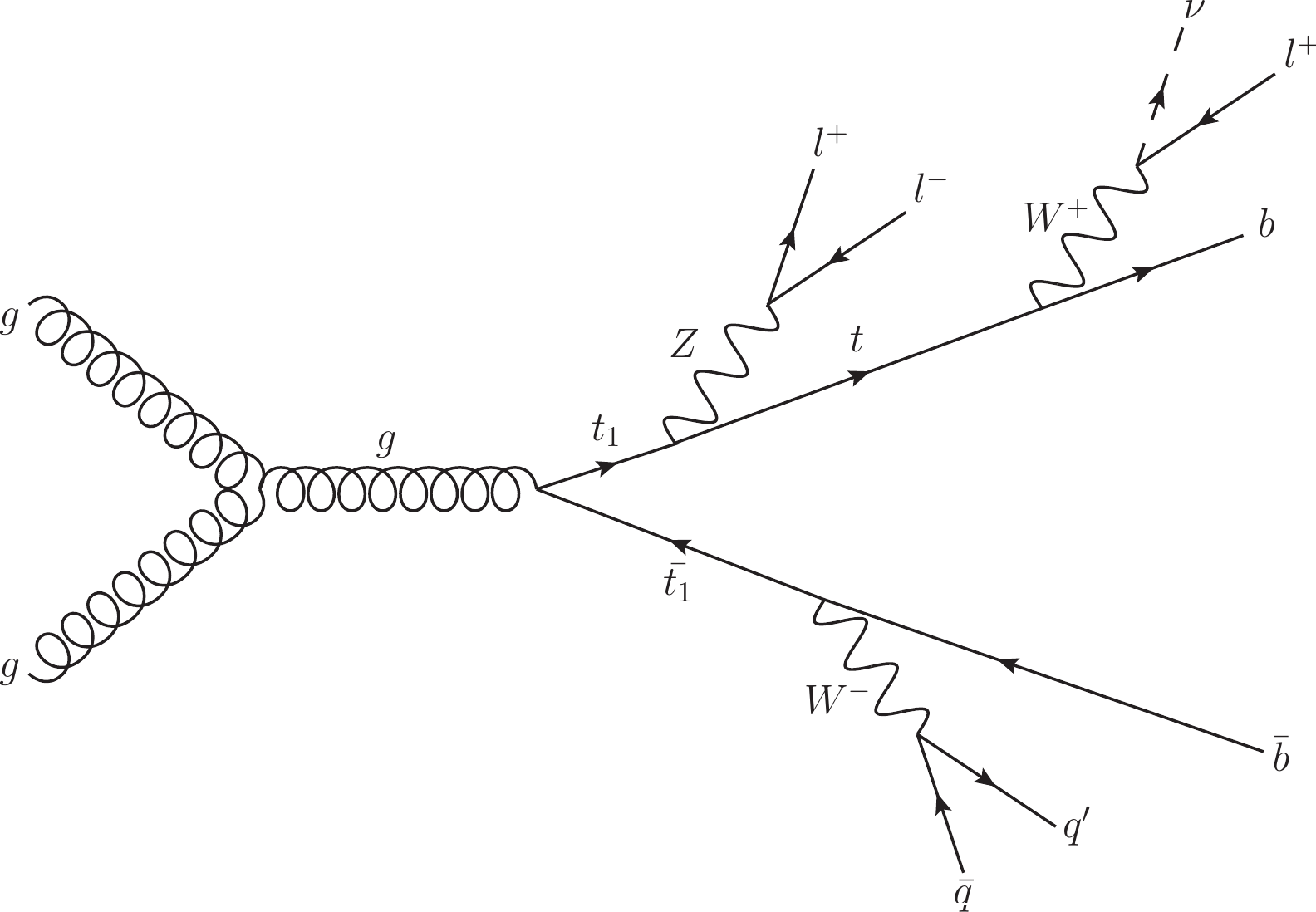, width=0.45\textwidth }
  	{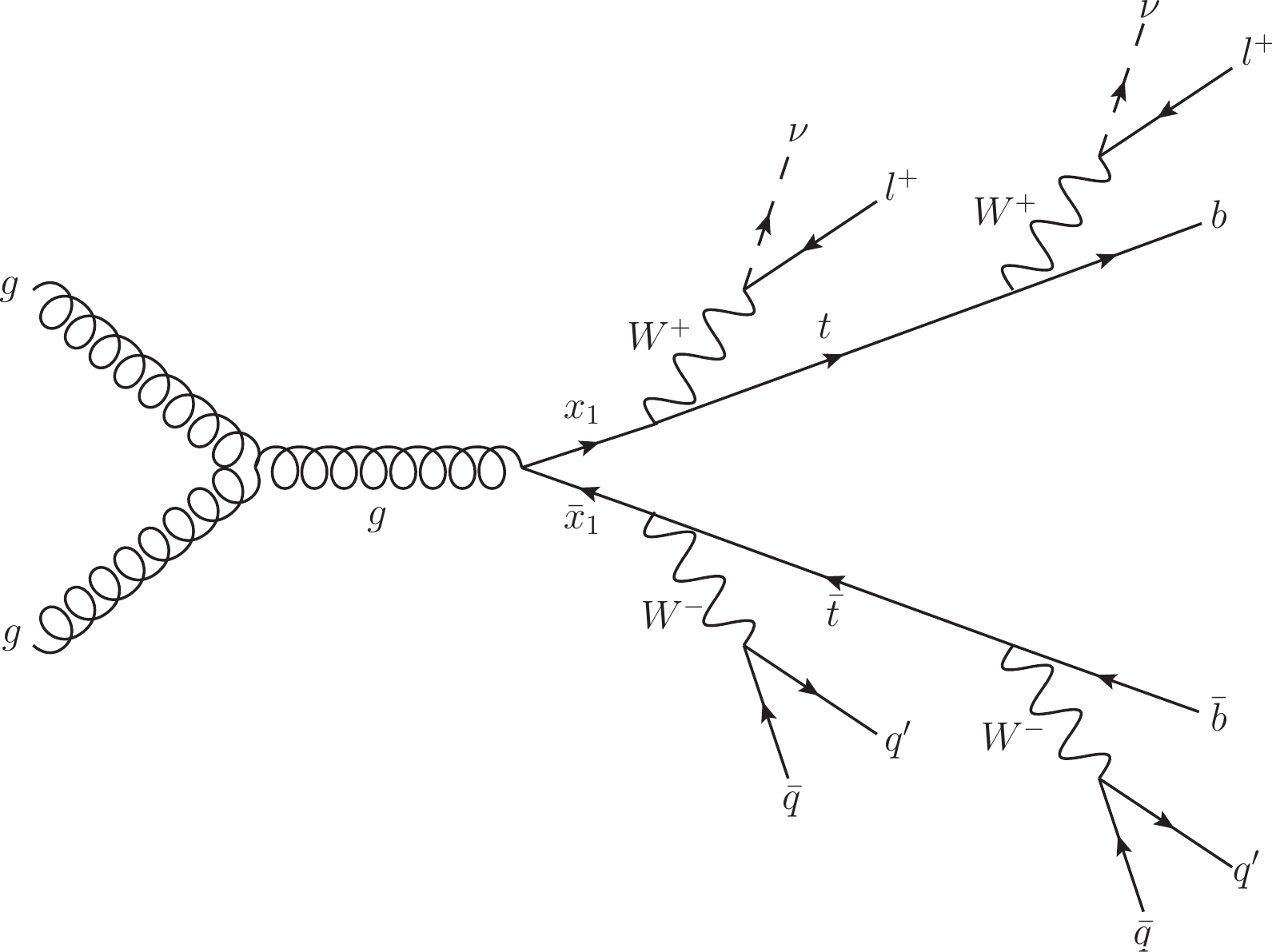, width=0.45\textwidth }
  	{Example Feynman diagram for $t_1$ pair production with a possible decay chain. \label{fig:t1_feynman}}
  	{Similar diagram for $x_1$ pair production.\label{fig:x1_feynman}}
	
	\paragraph{Dominant decay modes.}
	The process 
	$gg \ \to q \bar{q}$ plays the dominant role in the production of heavy top partners at the LHC. Therefore, we
	analyse the decay chains that start from pair produced quarks. 
	Figures~\ref{fig:t1_feynman} and~\ref{fig:x1_feynman}
	show two Feynman diagrams 
	for a possible decay chain of a $t_1\bar{t}_1$ and a $x_1\bar{x}_1$ pair. 
	
	In all points of the parameter space 
	that satisfy the selection criteria of section~\ref{sec:scans}, 
	the two lightest new quarks are $x_1$ and $t_1$. 
	The signatures from this model that could be observed early at the LHC will be therefore dominated by the decay modes of these 
	two quarks. 
	We also find that their mass difference 
	is always too small for the heavier of the two to decay into the lighter.
	Consequently, only the following channels are accessible for the decay of the two new quarks
	 \begin{equation}
	 	t_1 \to t Z \qquad t_1 \to t h \qquad t_1 \to b W^{+} \qquad x_1 \to t W^{+} \;.
		\label{eq:dominant_t1_x1_dec}
	 \end{equation}
	The lightest bottom-like quark $b_1$, which we always find to 
	be heavier than $t_1$ and $x_1$, decays predominantly via
	 \begin{equation}
	 	b_1 \to t W^{-} \;.
		\label{eq:dominant_b1_dec}
	 \end{equation}
	 The other possible decays,
	 $b_1 \to bZ$ and $b_1 \to b h$, 
	 are strongly suppressed because of the small off-diagonal couplings. 
	 Such small mixing is a consequence of the fact that the bottom quark is mainly
	 fundamental.
	 We find that the decay $b_1 \to t_1 W^{-}$ is not kinematically accessible.
	 
	\paragraph{Phenomenology of the $\mathbf{4}$ of $\mathbf{SO(4)}$.} 
	We consider
	$x_1$, $t_1$, $t_2$ and $b_1$ to form a $\mathbf{4}$ of $SO(4)$ when
	the maximal mass difference among the quarks is~$\lesssim 60$~GeV. 
	In this way 
	none of the new quarks can decay into another one, since
	decays through the $W$, $Z$ and $h$ bosons are not kinematically allowed. Consequently, 
	all these four new quarks can only decay to the SM top and bottom quarks. 
	 
	\paragraph{Phenomenology of the XX.}
	The phenomenology can be much richer if both charge 5/3 quarks are below 500~GeV and 
	no restriction on the maximal mass difference among the new quarks is imposed. 
	However, the exclusion limits from the CDF experiment in combination with 
	the upper bound of~500 GeV for early detection 
	imposes strong restrictions on the cascade decays that are kinematically allowed. 
	Often, the mass differences of these quarks are such that they only decay via the channels given in 
	\eqref{eq:dominant_t1_x1_dec} and \eqref{eq:dominant_b1_dec}. 
	The two lightest quarks are $x_1$ and $t_1$, where either of the two can be the lighter one. 
	Going up in mass, we find $b_1$ and $t_2$, or $t_2$ and $b_1$. The next heavier 
	quark is either $x_2$ followed by $t_3$, or vice versa. 
	The most common hierarchy is 
	\begin{equation}
		m_{x_1} < m_{t_1} < m_{b_1} < m_{t_2} < m_{x_2} \;. 
	\end{equation}
	A rarer mass pattern is 
	\begin{equation}
		m_{t_1} < m_{x_1} < m_{t_2} < m_{b_1} < m_{t_3} < m_{x_2} < m_{t_4} \;.
	\end{equation}
	The quarks that do not appear in these relations have masses above~500 GeV. 
	An example for a cascade decay accessible for various points is
	\begin{equation}
		t_2 \to x_1 W^{-} \to t W^{+} W^{-} \to b W^{+} W^{+} W^{-} \;.
		\label{eq:t2_cascade}
	\end{equation}
	Both this cascade decay and the dominant decay modes from eqs.~(\ref{eq:dominant_t1_x1_dec}) and~(\ref{eq:dominant_b1_dec})
	suggest that the model can easily yield multi-lepton final states plus many jets. 
	The SM is expected to produce only few events with such a signature. For this reason, we
	focus our study on final states with at least two leptons and multiple jets. 


\section{Model implementation in MadGraph and categorization of points}
\label{sec:implementation_and_grouping}

\subsection{Implementation}
\label{sec:implementation}

	We implement the model in MadGraph/MadEvent 4 (MG/ME) 
	\cite{MG-ME}. MG/ME is a matrix-element based 
	tree-level event generator that is capable of generating amplitudes and events 
	for any given model describing high energy physics interactions. 
	For such an event generator to be able to cope with a new physics model, the couplings
	and interactions of the new particles as defined in section \ref{sec:fermionic_sector}, in addition to the (modified)
	Standard Model interactions, have to be translated into a specific form. In MG/ME these couplings are 
	defined according to the
	convention from HELAS \cite{Murayama:1992gi}.
	The implementation of these couplings
	is done by means of the \verb+usermod v1+ framework. 
	The decay widths and branching ratios of 
	all unstable particles are calculated with BRIDGE \cite{BRIDGE}.
	
	We implement the model taking into account not only the couplings of the 
	newly introduced particles, but also the changes in the Standard Model couplings arising from 
	Higgs compositeness and from the mixing of the SM quarks with the new states.

\subsection{Benchmark points in the composite Higgs model parameter space}
\label{sec:grouping}

	Since the scan over the parameter space 
	was optimized to search for points that satisfy the selection criteria of section~\ref{sec:scans}, 
	the points returned are not necessarily very different from each other. For this reason, 
	we arrange the points in groups that are expected to exhibit a similar phenomenology and focus on the
	representatives of these groups for a detailed study. 
	We assign two points to the same group if
	all branching ratios of the new quarks with a mass below~500 GeV are of similar magnitude.
	When a group contains more than one point, 
	we use the mass of the lightest new quark $m_{q,\mathrm{low}}$ to select two representatives:
	the point with the lowest  value of $m_{q,\mathrm{low}}$ and the one with largest value of $m_{q,\mathrm{low}}$.
	In the following, these two 
	representatives will be referred to as low benchmark point (lBP) and high benchmark point (hBP) of a group. 
	For the discussion of the discovery potential, we will restrict 
	ourselves to the 30 benchmark points obtained in this way.

\section{Event generation and detector simulation}
\label{sec:Simulation_and_Reconstruction}

\subsection{Event generation}

	\TABLE{\begin{tabular}{rccc|rccc}
	BP  	& signature	        & $m_{q,\mathrm{low}}$ 	 &  cross section 	& 	 	      	&		  &         &   \\
		   	&			        & (GeV)		     		& (pb)				&		      	&			  & 		& 		\\
	\hline	                                     
	lBP 1	&XX			        & $x_1$: 365.6 	 		& 4.01	& lBP 2		&XX 		      &$x_1$: 366.0    &  3.85\\
	hBP 1	&XX, $\mathbf{4}$   & $x_1$: 429.8 	 		& 2.12	& hBP 2		&XX, $\mathbf{4}$ &$x_1$: 408.9    &  2.50\\	
	lBP 3	&XX, $\mathbf{4}$ 	& $x_1$: 366.1 	 		& 7.90	& lBP 4		&XX, $\mathbf{4}$ &$x_1$: 400.9    &  4.45   \\
	hBP 3	&XX, $\mathbf{4}$   & $x_1$: 404.0 	 		& 3.94	& hBP 4		&XX, $\mathbf{4}$ &$x_1$: 462.9    &  1.61	\\	
	BP 5	&XX			        & $x_1$: 366.7 	 		& 3.45	& BP 6		&XX, $\mathbf{4}$ &$x_1$: 463.9    &  1.57	\\
	BP 7	&XX, $\mathbf{4}$   & $x_1$: 461.5 	 		& 1.70	& BP 8		&XX, $\mathbf{4}$ &$x_1$: 456.9    &  1.58	\\
	BP 9	&XX			        & $x_1$: 365.7 	 		& 3.88	& BP 10	 	&XX, $\mathbf{4}$ &$t_1$: 316.6    &  10.78	\\
	lBP 11	&XX, $\mathbf{4}$   & $x_1$: 377.5 	 		& 6.20	& lBP 12	&XX 		      &$x_1$: 367.7    &  4.29	\\
	hBP 11	&XX, $\mathbf{4}$   & $t_1$: 393.3 	 		& 3.86	& hBP 12	&XX, $\mathbf{4}$ &$x_1$: 391.5    &  5.52   \\	
	BP 13	&XX, $\mathbf{4}$   & $x_1$: 373.5 	 		& 7.80	& BP 14	 	&XX, $\mathbf{4}$ &$t_1$: 343.4    &  7.91	\\
	lBP 15	&$\mathbf{4}$		& $x_1$: 365.8 	 		& 5.58	& lBP 16	&$\mathbf{4}$     &$x_1$: 365.4    &  5.18	\\
	hBP 15	&$\mathbf{4}$		& $x_1$: 376.7 	 		& 4.40	& hBP 16	&$\mathbf{4}$     &$x_1$: 450.2    &  1.43  \\
	lBP 17	&$\mathbf{4}$		& $x_1$: 365.9 	 		& 5.23	& lBP 18	&$\mathbf{4}$     &$x_1$: 365.2    &  5.52	\\
	hBP 17	&$\mathbf{4}$		& $x_1$: 438.3 	 		& 1.66	& hBP 18	&$\mathbf{4}$     &$x_1$: 414.2    &  2.53  \\
	BP 19	&$\mathbf{4}$		& $t_1$: 375.3 	 		& 4.92	& BP 20		&$\mathbf{4}$     &$t_1$: 379.8    &  4.35	\\
	\end{tabular}
	\caption{
	The 30 benchmark points of the two multiplet model with the mass of the lightest 
	quark $m_{q,\mathrm{low}}$ and the total cross section for pair production of all quarks below 500~GeV. 
	The cross sections are calculated at leading order for $\sqrt{s}=7$~TeV.
	\label{table:signal_xsections}}}
	
		For each benchmark point, we produce $10^5$ signal events with MG/ME. 
	In particular, we generate events for pair production of all new quarks that have a mass
	below 500 GeV. The outcome of the MG/ME event generation is a Les Houches event file~\cite{LHE_2006}, 
	which we process with Pythia 6~\cite{Sjostrand:2006za} for the showering 
	and hadronization of the partonic events and for the simulation of the underlying event. 
	Table~\ref{table:signal_xsections} lists the mass of the 
	lightest particle $m_{q,\mathrm{low}}$ and the total leading order cross section for pair production of all considered quarks for each 
	benchmark point.
	
	As already mentioned in section~\ref{sec:phenomenology}, we focus on final state signatures with 
	at least two charged leptons and multiple jets. Consequently, every SM process that can lead to such 
	final states represents a possible background. Table~\ref{table:sm_backgrounds} lists the leading order 
	cross section and the number of generated events for all relevant background processes. 
	Note that single top
	production was neglected for this study. Its contribution is 
	expected to be within the uncertainty of the pair production cross section.
	In order to estimate correctly the momentum spectrum of the jets in the transverse plane of the detector, 
	we generate all partonic multiplicities needed for the SM backgrounds in MG/ME and use Pythia 
	for the parton shower. The overlap between the phase-space description of the matrix-element
	calculator and the parton shower is removed using the MLM parton-jet matching prescription~\cite{Alwall:2007fs}.
	For the signal samples, 
	the jets produced by the parton shower in the decay of
	very heavy particles are known to be satisfactory~\cite{Plehn:2005cq}. 
	The underlying event is simulated with Pythia. For all samples, we use the parton distribution function set CTEQ6L1.	
	
	\TABLE{\begin{tabular}{llrr}
	process	&			& 	cross section (pb) 	& \# of generated events \\
	\hline
	$Z $                &$ + \le 3 \ \mathrm{j}$	    &	2400	 		&	$1302405^{*}$			\\
	$W $                & $ + \le 3 \ \mathrm{j}$	    &	24170			&	$12270142^{*}$		\\
	$VV $               &$ + \le 1 \ \mathrm{j}$	    &	4.8				&	$113764^{*}$			\\
	$W^{\pm}W^{\pm}$    &$ + 2 \ \mathrm{j}$		    &	0.2119			&	$47070\ $				\\
	$W^{+}W^{-}W^{\pm}$ &$ $		                    &	0.04105			&	$49999\ $				\\
	$t\bar{t} $         &$ + \le 3 \ \mathrm{j}$	    &	95				&	$1395630^{*}$			\\
	$t\bar{t}W^{\pm} $  &$ + \le 1 \ \mathrm{j}$	    &	0.1687			&	$66266\ $				\\
	$t\bar{t}Z      $   &$ $		                    &	0.1038			&	$49999\ $				\\
	\end{tabular}
	\caption{Background processes with the corresponding cross section 
	and the number of generated events. 
	The di-vector boson sample $VV+\mathrm{jets}$ includes 
	all processes with two $W$ or $Z$ bosons, except for the case of two $W$ bosons with the same charge.
	In the first three samples, the vector bosons are forced to decay leptonically.
	The $(^{*})$ indicates that we used the MG/ME Les Houches events made 
	available from the LCG Monte-Carlo Data Base~\cite{Belov2008222}
	instead of generating the events ourselves.
	\label{table:sm_backgrounds}}}

	We would like to point out that the samples for the background processes were generated within the SM. 
	We did not take into account the changes of the SM couplings introduced in the composite Higgs model. 
	These modifications differ for each point in the parameter space of the model, but we expect the resulting 
	effects on the SM backgrounds to be small. 
	Also note that we only consider pair production of the new quarks for the signal samples. We neglect the 
	contributions of other processes (such as single quark production) to the signal yield in multi-lepton final states. 
	These additional contributions to the signal would enhance the excess over the SM expectation.

\subsection{Detector simulation}
\label{sec:delphes}
	We use DELPHES~\cite{Ovyn:2009tx} for the simulation of the response of a typical LHC detector. 
	DELPHES is a recently developed 
	simulation framework for a generic collider experiment. As CMS is one of the 
	two general purpose detectors at the LHC, we use the CMS detector card for the
	DELPHES detector simulation. We reconstruct the jets with the anti-$k_t$ jet-clustering
	algorithm~\cite{Cacciari:2008gp} and use a cone radius 
	$\Delta R=\sqrt{\Delta \phi^{2}+\Delta \eta^{2}}=0.5$. 
	$\phi$ denotes the azimuthal angle and the 
	pseudorapidity $\eta$ is defined as $\eta= - \ln \tan \frac{\theta}{2}$, where $\theta$ is the angle 
	between the beam pipe and the trajectory of the particle.
	To adapt the performance of DELPHES to our needs, we make the 
	following modifications.
	\begin{itemize}
		\item 	In DELPHES, the possibility of a jet being reconstructed as an electron is
				not taken into account. This, however, is expected to be a relevant source of fake electrons. 
				In ref.~\cite{Acosta:922757}, the probability
				for a jet to be reconstructed as an isolated, identified electron is estimated to be at a level of $6\cdot 10^{-6}$.
				We use this result and add jets to the isolated electron collection with the stated global probability.
		\item 	We set the global tracking efficiency to 100\% for tracks with a transverse momentum of at least 0.9~GeV, 
				but remove electrons from the electron 
				collection with a probability of 10\%. 
	\end{itemize}
	
\subsection{Lepton and jet identification}
\label{sec:lepton_jet_id}
	We outline a robust and simple event
	selection that is suitable for early data from the LHC. 

	\paragraph{Charged lepton selection.} For the electrons and the muons, we demand a transverse momentum 
	$p_{T} > 20 \ \mathrm{GeV}$ and a pseudorapidity $|\eta|<2.4$. The first cut ensures a robust identification
	of electrons and muons, both offline and on trigger level, whereas the second cut is made to restrict the leptons to 
	the volume of the tracker. For this study, we
	are interested in prompt leptons coming from vector boson decays. 
	To discriminate against leptons coming from semileptonic hadron decays, we apply a relative isolation. 
	In particular, we sum the $p_{T}$ of the tracks in a cone with $\Delta R<0.3$ 
	around the electron or muon under analysis and require this value to be smaller than 5\% of the lepton momentum.

	\paragraph{Jet selection.} To obtain a robust jet selection, we demand the $p_{T}$ of a jet to be larger than $50\ \mathrm{GeV}$
	and require $|\eta|<3$. The conservative choice of $p_{T}>50$~GeV
	is made to minimize the contribution of fake jets. 
	The second cut marks the end of the electromagnetic and 
	hadronic calorimeters.
	As electrons may be reconstructed as possible jet candidates, we reject those jets that are matched within
	$\Delta R<0.2$ to an isolated electron. The jet collection can be further cleaned from such electrons by 
	requiring that the jets
	should have an electromagnetic fraction (electromagnetic over hadronic energy deposits) of less than 0.98.

	\paragraph{Purity and efficiency of the lepton selection.} Imposing a harder cut on the lepton isolation enhances the purity of
	the selection but causes the efficiency to decrease. The goal is to achieve a pure selection of prompt leptons 
	without losing too much efficiency. By purity we define the number of isolated leptons matched
	to prompt MC leptons divided by the number of isolated leptons. The efficiency is defined as the number of isolated 
	leptons divided by the number of MC prompt leptons. The number of matched isolated leptons is obtained by counting
	the ones that satisfy both criteria: 
	\begin{itemize}
		\item they have a prompt MC lepton within a cone of $\Delta R<0.2$
		\item the equation $\frac{|p_{T, iso}-p_{T, MC}|}{p_{T, MC}}<0.2$ holds.
	\end{itemize}
	For the $t\bar{t}$ sample from table~\ref{table:sm_backgrounds}, we find an efficiency of 
	83\% and a purity of 97\% for the electrons. 
	For the muons, we obtain an efficiency of 91\% and a purity of 99\%.

\section{Discovery potential at the LHC}	
\label{sec:discovery}
	\subsection{Identification of promising channels}
	\label{sec:channel_selection}
	
	\DOUBLEFIGURE{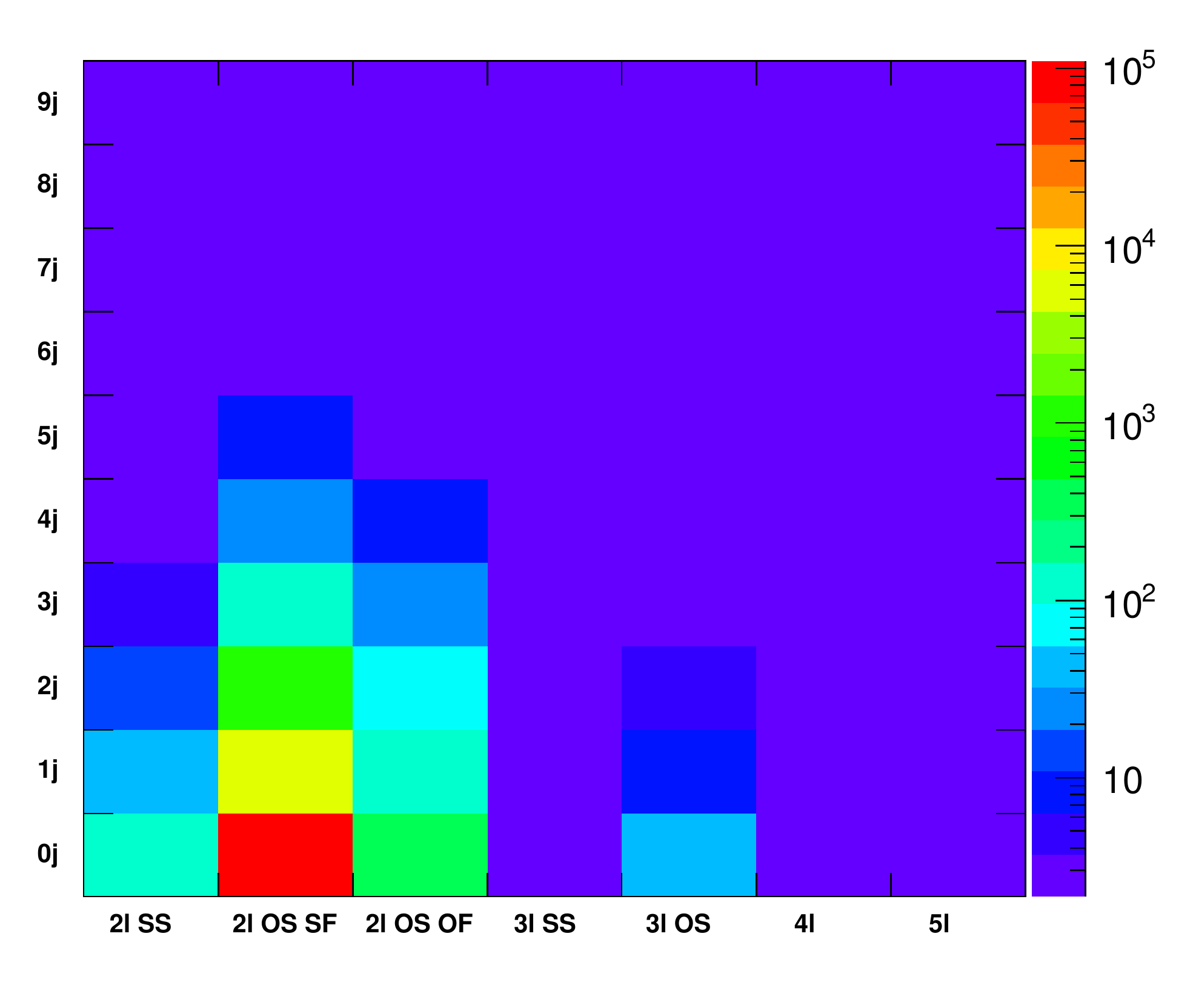,width=0.5\textwidth}
	{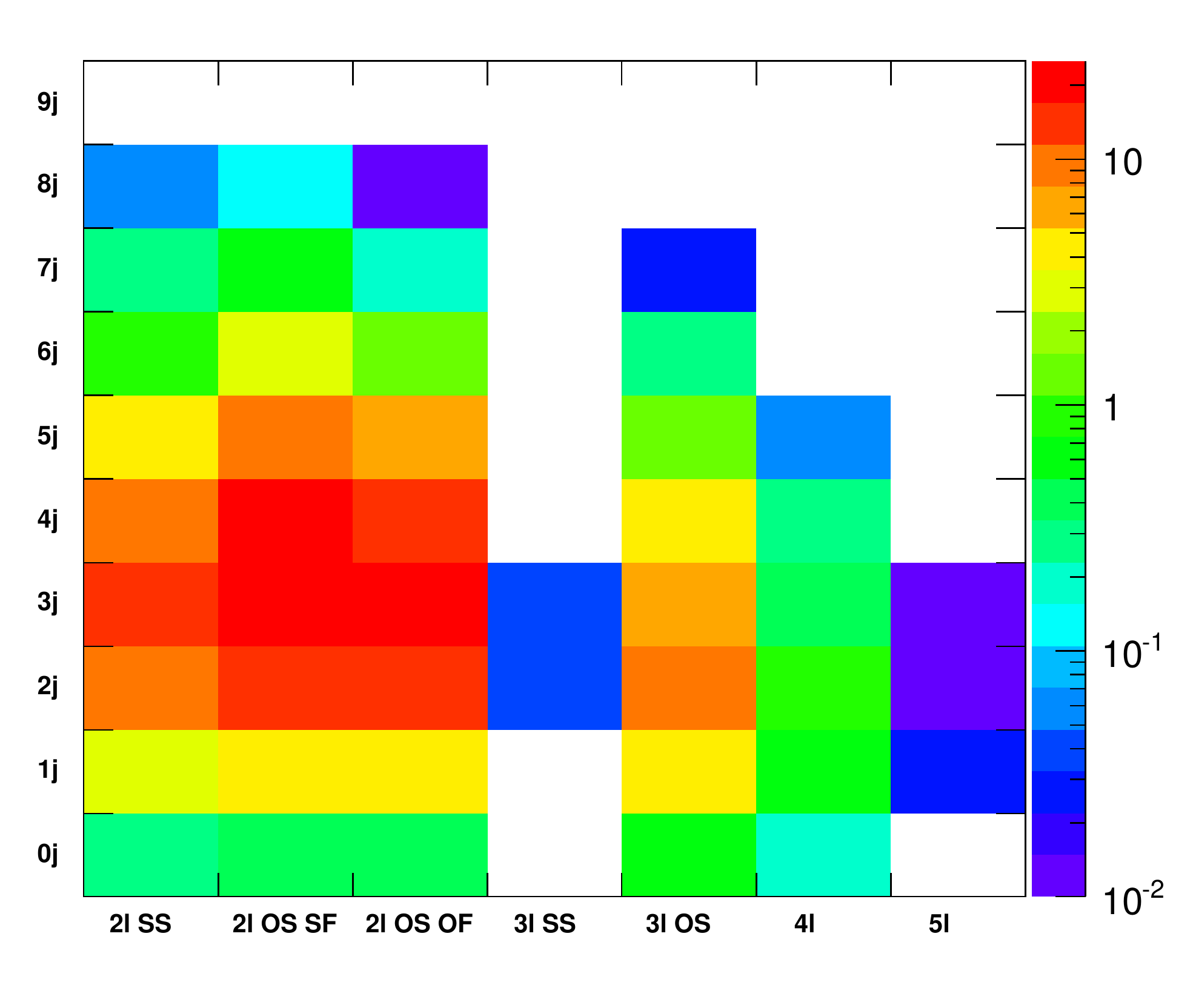,width=0.5\textwidth}
	{The upper limit for the total number of background events based on a 95\% confidence level. 
	This plot includes all relevant backgrounds scaled for 
	an integrated luminosity of $200\ \mathrm{pb}^{-1}$. The y-axis shows the jet multiplicity, 
	whereas the lepton configuration is given on the x-axis. \label{figure:total_bg_upper}}
	{The same plot as on the left hand side, but showing the lower limit for the expected number of signal events for 
	BP~10.\label{figure:lower_limit_group10}}	

	After applying the lepton and jet selection defined in section~\ref{sec:lepton_jet_id}, 
	we investigate the number of events for a given jet multiplicity and lepton configuration for
	each background and signal process. The lepton configurations go from 
	di-lepton events - same-sign (SS) or opposite-sign (OS) - to events with up to five charged
	leptons in the final state. Each configuration, which is characterized by a certain lepton combination and jet multiplicity, 
	is interpreted as a specific signal region with an associated cut efficiency. Since these cut efficiencies are
	based on a finite MC statistics, we observe certain 
	configurations with non-vanishing signal but zero background events. 
	To avoid this issue, we 
	calculate for all configurations lower and upper limits for the cut efficiency for the signals and backgrounds respectively, 
	based on a 95\% confidence level. This can be interpreted as a worst case scenario for the 
	discovery of the model. The expected number of signal and background events are obtained by mul- 
   	\EPSFIGURE{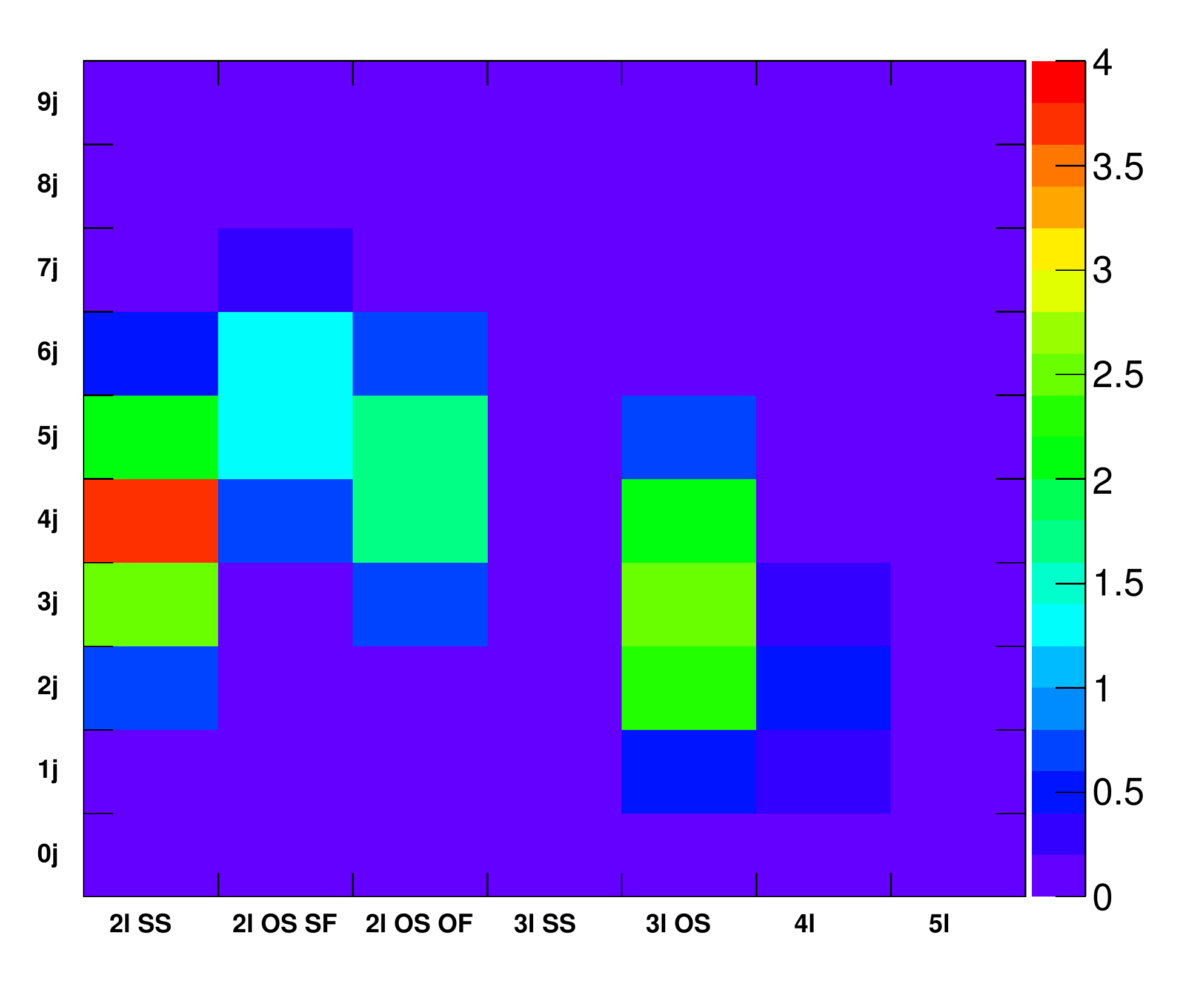, width=0.5\textwidth}
   	{The expected number of signal events for BP~10 divided by 
   	the total number of events from the SM background. \label{figure:group10_ratio}}	
	\hspace{-24pt} tiplying these efficiencies by the integrated
	luminosity of $200\ \mathrm{pb}^{-1}$
	times the corresponding cross section (as listed in tables~\ref{table:signal_xsections} and~\ref{table:sm_backgrounds}).	

	In figures~\ref{figure:total_bg_upper} and~\ref{figure:lower_limit_group10} we plot the
	jet multiplicity versus the lepton configuration, respectively for the total SM background and for the signal for BP~10.
	We denote by SS the configurations in which all the leptons have the same charge. 
	Configurations in which at least one lepton has a different charge are denoted by OS. 
	In the OS di-lepton case, we also distinguish between the opposite-flavor (OF) and same-flavor (SF) configurations.
	For the bins in figure~\ref{figure:total_bg_upper} for which 
	zero MC background events were found, we calculate 
	an upper limit of 2.13 background events with a confidence of 95\%.
	This number is dominated by the contributions from $W+\mathrm{jets}$ and $Z+\mathrm{jets}$
	due to their large cross sections and the limited MC statistics.  
	In figure~\ref{figure:group10_ratio}, we plot the 
	number of signal events for 
	BP~10 divided by the total number of background events. In terms of number of expected 
	events over the SM background, we can see that the final states with SS di-leptons and OS tri-leptons are the most promising 
	channels for a possible discovery with $200 \ \mathrm{pb}^{-1}$ of collision data. 
	This observation holds for all the 30 benchmark points. 
	The decrease in the 
	plotted S/B ratio for large jet multiplicities
	is not expected in collision data.
	This effect is due to the combination of a finite number of MC events with
	the calculation of an upper limit on the number of background events. 
	In the light of the above discussion, 
	we will focus on the four channels:
	SS di-lepton with 3 or 4 jets and OS tri-lepton with 2 or 3 jets.

	\subsection{Inclusive discovery potential}
	\label{sec:disc_pot}
	In order to quantify the discovery reach 
	in the four channels above, we calculate the
	probability for the expected signal + background observation to be caused by a fluctuation in the background 
	distribution. We use 
	$2\log X$ as a test statistic, where $X$ is the ratio of the likelihood function for the signal + background hypothesis $H_{1}$
	to the likelihood function for the background hypothesis $H_{0}$ \cite{Read,Junk1999435}.	

	The likelihood ratio $X_{i}$ for the channel $i$
	can be defined as 
	\begin{equation}
		X_{i}=\frac{\mathcal{L}_{H_{1}, i}}{\mathcal{L}_{H_{0}, i}}\;,
	\end{equation}
	where 
	\begin{equation}
		\mathcal{L}_{H_{1},i}=\frac{e^{-(s_{i}+b_{i})}(s_{i}+b_{i})^{d_{i}}}{d_{i}!} 
		\quad , \quad
		\mathcal{L}_{H_{0},i}=\frac{e^{-(b_{i})}(b_{i})^{d_{i}}}{d_{i}!}\;.
	\end{equation}
	Here, $s_{i}$ and $b_{i}$ denote the number of signal and background events, respectively, and $d_{i}$ is the 
	number of observed candidates. Since the statistic $2\log X$ for the outcome of multiple channels is the sum of the
	test statistics of the channels separately, we use $2 \sum_{i=1}^{4} \log X_{i}$ for the combined four channels defined in 
	section~\ref{sec:channel_selection}. We define the confidence level as 
	\begin{equation}
		\mathrm{CL_b}=P_b(X<X_{\mathrm{obs}})\;,
	\end{equation}
	where the probability sum assumes the presence of the background only.
	Note that the background confidence $1-\mathrm{CL_{b}}$ 
	expresses the compatibility of the observation with the background hypothesis, since $\mathrm{CL_{b}}$ is 
	the probability that the background processes would give fewer than or equal to the number of events observed. 
	For this reason, we use $\mathrm{CL_b}$ to quantify the discovery potential.
	The background confidence $1-\mathrm{CL_{b}}$ can be compared
	with the widely used notion of standard deviations ($\sigma$)
	by using the convention from ref.~\cite{PDG}\footnote{See e.g. table 32.1}. 
	There, a 3$\sigma$ and 5$\sigma$ excess beyond the 
	background expectation corresponds to a one-sided background confidence level of 
	$1-\mathrm{CL_{b}}=1.35\cdot 10^{-3}$ and $1-\mathrm{CL_{b}}=2.87 \cdot 10^{-7}$ respectively.

	The distribution of the test statistic for $H_{0}$ and $H_{1}$, 
	often referred to as the test statistic probability density function (tPDF), are obtained by throwing Poisson 
	numbers around $s_{i}+b_{i}$ and $b_{i}$ as a replacement for $d_{i}$. 
	The confidence level $\mathrm{CL_b}$ and its uncertainty is calculated as follows. 
	In the presence of data, $\mathrm{CL_b}$ is given by the integral of the tPDF of the background hypothesis from 
	$-\infty$ to the measured value of $2\log X$. For this study, 
	we replace this value by the mean of the tPDF for the signal + background hypothesis to substitute collision data. 
	The uncertainty on $\mathrm{CL_b}$ is then obtained by changing the integration limit to the mean plus/minus one standard deviation of the 
	signal + background tPDF. To claim a 5$\sigma$ excess over the background expectation, 
	we have to be sensitive to $\mathrm{CL_{b}}$ at the order of $10^{-7}$. For this reason, we generate 
	$10^{9}$ pseudo-experiments for each of the tPDFs for $H_{0}$ and $H_{1}$.

	In table~\ref{table:bg_each_channel} we list the expected number of
	events for all background processes for the four channels considered. The corresponding values for the
	30 signal benchmark points, including the results for the confidence level 
	$\mathrm{CL_{b}}$,
	are given in table~\ref{table:CLb} of appendix~\ref{sec:CLb_table_appendix}.  
	Even in the worst case scenario, 
	we expect a signal evidence of at least 3$\sigma$ for 23 benchmark points. 
	For 10 points among these 23, the central $\mathrm{CL_b}$ value
	corresponds to an excess over the SM expectation of at least 5$\sigma$.

	\TABLE[t]{\begin{tabular}{ccccc}
	process						&	2l SS + 3j 			&	2l SS + 4j			& 3l OS + 2j  	& 3l OS + 3j		\\
	\hline                  	
	$Z+\mathrm{jets}$			&	1.7					&	1.0					&	1.7			&  1.7		      \\
	$W+\mathrm{jets}$			&	1.8					&	1.1					&	1.1			&  1.1		      \\
	$VV+\mathrm{jets}$			&	0.075				&	0.023				&	0.49		&  0.12		      \\
	$W^{\pm}W^{\pm}jj$			&	0.099				&	0.019				&	0.0024		&  0.0024		      \\
	$W^{+}W^{-}W^{\pm}$			&	0.0035				&	0.00044			 	&	0.0012		&  0.00044	   	   \\
	$t\bar{t}+\mathrm{jets}$	&	2.1					&	0.83				&	0.52		&  0.25		      \\
	$t\bar{t}W^{\pm}j$			&	0.19				&	0.075				&	0.052		&  0.016		      \\
	$t\bar{t}Z^{\pm}$			&	0.036				&	0.011				&	0.085		&  0.063		      \\
	\hline                  	                                                                               	   
	total 						&	5.90				&	3.02				&	3.86		&  3.16			\\
	\end{tabular}
	\caption{The upper limit on the number of expected events for $200\ \mathrm{pb}^{-1}$ of data for each of the
	background processes at a confidence level of 95\%. 
	Systematic uncertainties on cross sections are not taken into account. 
	\label{table:bg_each_channel}}}

	\subsection{Discovery potential of two benchmark points}
	\label{sec:two_smoking_guns}

	We now focus on the 
	discovery potential of two promising benchmark points.
	One is BP~10, which has both $x_1$ and $x_2$ below 500~GeV; the other is lBP~18.
	Both benchmark points exhibit a $\mathbf{4}$ of $SO(4)$ and
	have large cross sections (10.78~pb and 5.52~pb 
	respectively), yielding a relevant excess over the SM background.
	We use a simple cut-based analysis and
	outline some features of their specific phenomenology.

	\paragraph{Phenomenology of the two benchmark points.} As we can see from table~\ref{table:signal_xsections},
		the lightest new quark for BP~10 is the top-like $t_1$ with a mass of 316.6~GeV. 
		The full mass hierarchy for the new quarks with a
		mass below 500~GeV reads
		\begin{eqnarray*}
			m_{t_1}\ (316.6) &<& m_{x_1}\ (365.2) < m_{t_2}\ (374.4) < m_{b_1} \ (377.3) \\
			                    &<& m_{t_3}\ (377.9) < m_{x_2}\ (395.3) < m_{t_4} \ (473.3)\;, 
		\end{eqnarray*}
		where the masses are given in GeV. For this point, the mass difference between the $t_4$ quark
		and the other quarks is such as to allow the $t_4$ to decay into most of them. The full list of branching
		ratios for all the above listed quarks can be seen in table~\ref{table:group_10_br}.

		For lBP~18, the mass hierarchy is 
		\begin{eqnarray*}
			m_{x_1}\ (365.2) < m_{t_1}\ (367.6)  < m_{b_1} \ (373.9) < m_{t_2}\ (403.1)\;.  
		\end{eqnarray*}	
		Given that the maximal mass difference among these quarks is about 40~GeV, their decay modes are described 
		by eqs.~\eqref{eq:dominant_t1_x1_dec} 
		and \eqref{eq:dominant_b1_dec}. In table~\ref{table:group_18_br} we list the branching ratios corresponding 
		to these decay modes. 

		\TABLE{\begin{tabular}{@{}l@{ }l|l@{ }l|l@{ }l@{}}
		BP~10: \\
		\hline
		BR($t_1\to b W^+$):   & $2.74\cdot 10^{-1}$     & BR($t_2\to b W^+$):  & $5.47\cdot 10^{-5}$    & BR($t_3\to b W^+$):  & $2.09\cdot 10^{-2}$   \\
		BR($t_1\to t Z $):    & $1.68\cdot 10^{-1}$     & BR($t_2\to t Z $):   & $3.80\cdot 10^{-1}$    & BR($t_3\to t Z $):   & $8.95\cdot 10^{-1}$   \\
		BR($t_1\to t h $):    & $5.58\cdot 10^{-1}$     & BR($t_2\to t h $):   & $6.20\cdot 10^{-1}$    & BR($t_3\to t h $):   & $8.17\cdot 10^{-2}$    \\
		\hline
		BR($t_4\to b W^+$):   & $6.89\cdot 10^{-2}$     & BR($t_4\to t Z$):    & $3.56\cdot 10^{-1}$    & BR($t_4\to t_3 Z$):  & $5.98\cdot 10^{-3}$ \\
    	BR($t_4\to b_1 W^+ $):& $1.17\cdot 10^{-2}$     & BR($t_4\to t_1 Z $): & $7.03\cdot 10^{-2}$    & BR($t_4\to t h $):   & $4.42\cdot 10^{-1}$ \\
		BR($t_4\to x_1 W^- $):& $3.01\cdot 10^{-2}$     & BR($t_4\to t_2 Z $): & $1.38\cdot 10^{-3}$    & BR($t_4\to t_1 h $): & $1.02\cdot 10^{-2}$ \\
		\hline
		BR($b_1\to t W^- $):  &  $9.96\cdot 10^{-1}$    & BR($b_1\to b Z $):   & $3.78\cdot 10^{-5}$    & BR($b_1\to b h $):   & $2.29\cdot 10^{-5}$  \\	
		\hline
		BR($x_1\to t W^+ $):  &  $1.00$                 & BR($x_2\to t W^+ $): & $9.97\cdot 10^{-1}$    &                   &           \\
		\hline
		\end{tabular}
		\caption{The branching ratios for the seven quarks with a mass below 500~GeV for BP~10.
		Note that the branching ratios may not add up to 1 as possible three-body decays might contribute.}
		\label{table:group_10_br}}			

		\TABLE{\begin{tabular}{@{}l@{ }l|l@{ }l|l@{ }l@{}}
		lBP~18: \\
		\hline
		BR($t_1\to b W^+$):& $1.56\cdot 10^{-2}$     & BR($t_2\to b W^+$):& $3.02\cdot 10^{-1}$   & BR($b_1\to t W^- $):& $1.00$   \\
		BR($t_1\to t Z $):& $9.48\cdot 10^{-1}$      & BR($t_2\to t Z $): & $1.40\cdot 10^{-1}$   & BR($b_1\to b Z $):  & $4.06\cdot 10^{-5}$   \\
		BR($t_1\to t h $):& $3.64\cdot 10^{-2}$      & BR($t_2\to t h $): & $5.58\cdot 10^{-1}$   & BR($b_1\to b h $):  & $2.47\cdot 10^{-5}$    \\
		\hline
		BR($x_1\to t W^+ $):& 1.00                   &                   &                       &           \\
		\hline
		\end{tabular}
		\caption{The branching ratios for the four quarks with a mass below 500~GeV for lBP~18.}
		\label{table:group_18_br}}

	\paragraph{Cut-based analysis.} 
		We outline a simple, cut-based 
		analysis for the two benchmark points
		to illustrate a complementary way to investigate the discovery potential of the model.
		For this analysis we use the lepton and jet selections defined in section~\ref{sec:lepton_jet_id}. 
		Given the results from tables~\ref{table:bg_each_channel} 
		and~\ref{table:CLb}, 
		we ask as preselection 
		to have at least two same-sign (isolated) leptons (e, $\mu$) with $p_{T} > 20$~GeV and $|\eta|<2.4$.
		
		\DOUBLEFIGURE{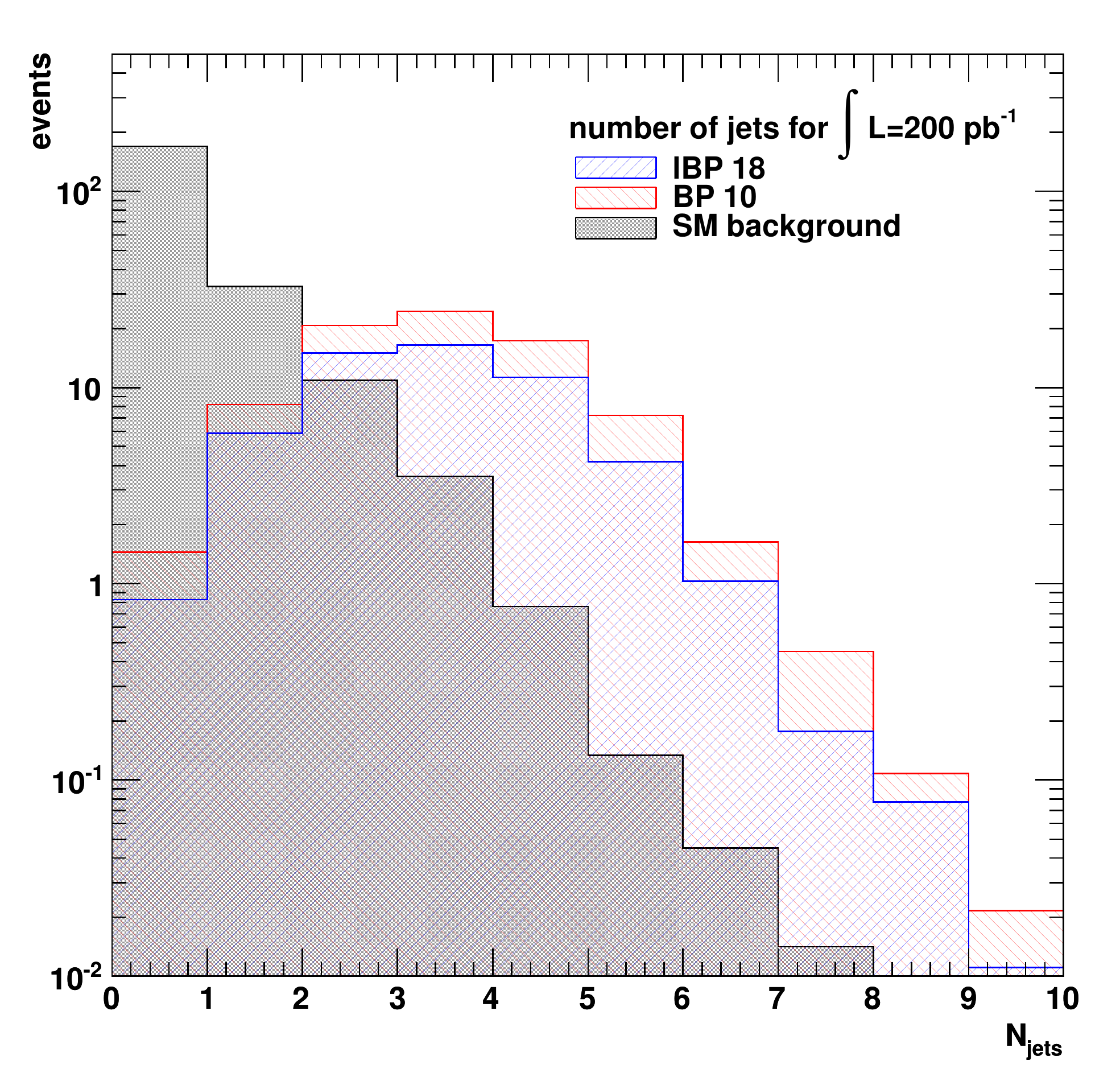,width=0.45\textwidth}
		{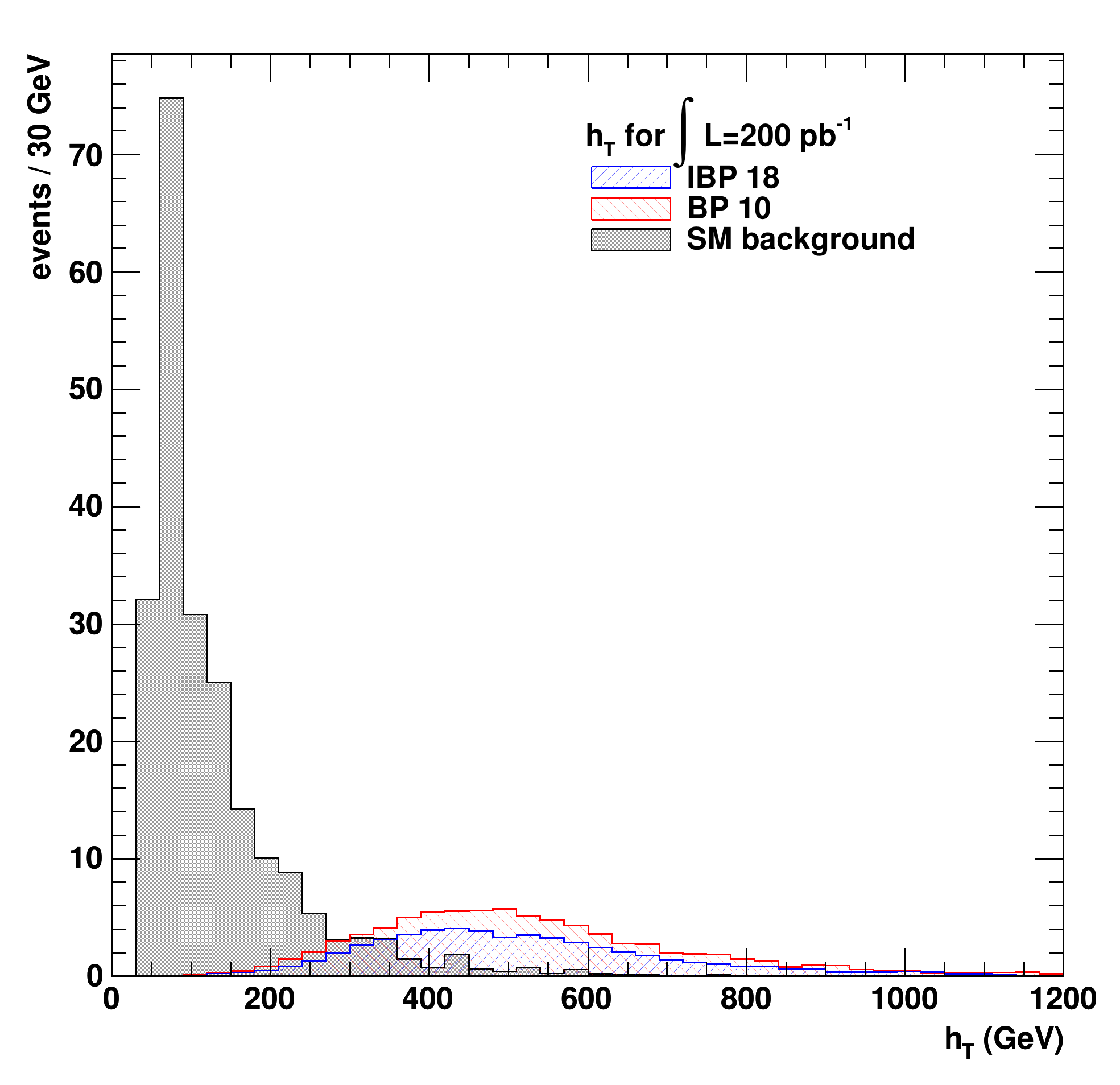,width=0.45\textwidth}
		{Number of jets per event for BP~10 (red), lBP~18 (blue) 
		and the SM background as from table~2 (gray).
		The plot is scaled to give the number of expected events for an integrated luminosity of 
		$200 \ \mathrm{pb}^{-1}$.\label{fig:n_jets}}
		{The scalar sum of the $p_{T}$ of all selected leptons and jets per event after preselection.
		The scaling and the color code is the same as on the left.\label{fig:ht}}
		
		Figure~\ref{fig:n_jets} shows the expected number of
		jets per event for BP~10, lBP~18  and the SM background after preselection 
		for an integrated luminosity of $200 \ \mathrm{pb}^{-1}$.
		Based on these distributions we 
		impose a cut on at least 2 jets, where the jets are requested to have $p_{T}> 50$~GeV.
		As a next step, we make use of the variable $h_{T}$,
		which is defined as the scalar sum of the transverse momentum of the selected jets and leptons per event.
		In figure~\ref{fig:ht} we show the overlaid distributions of $h_{T}$ for BP~10 (red), lBP~18 (blue)
		and the SM background (grey) scaled for $200 \ \mathrm{pb}^{-1}$ of data. The distributions were
		obtained only after imposing the preselection cut. Clearly, this variable can be used
		as a powerful cut to suppress the background contribution. For this reason, we require an 
		$h_{T} > 300$~GeV for the events to pass this cut.
		From figure \ref{fig:hardest_jet} we see that the signal distributions of the $p_T$ of the hardest jet peak at larger values
		than the corresponding SM background distribution. Consequently, we impose a cut at 90~GeV on this variable. 
		Summarizing, we impose the following cuts:
		\begin{enumerate}
			\item at least 2 jets with $p_T >  50$~GeV, 
			\item $h_T > 300$~GeV and
			\item $p_T$ of the leading jet $> 90$~GeV.
		\end{enumerate}
		
		\TABLE{\begin{tabular}{lccr@{}l}
		sample                       & preselection eff.  		  & total selection eff. 		&  \multicolumn{2}{c}{expected \# of events} \\
		\hline                                                   
		$Z+\mathrm{jets}$		     & $7.22\cdot 10^{-5}$        & $1.54\cdot 10^{-6}$       &    $\qquad 0.74  $&$^{+0.74}_{-0.37}$    \\
		$W+\mathrm{jets}$		     & $2.97\cdot 10^{-5}$        & $1.63\cdot 10^{-7}$       &    $\qquad 0.79  $&$^{+0.79}_{-0.39}$    \\
		$VV+\mathrm{jets}$		     & $2.19\cdot 10^{-2}$        & $6.15\cdot 10^{-4}$       &    $\qquad 0.59  $&$^{+0.07}_{-0.07}$    \\
		$W^{\pm}W^{\pm}jj$		     & $2.32\cdot 10^{-2}$        & $9.92\cdot 10^{-3}$       &    $\qquad 0.42  $&$^{+0.02}_{-0.02}$    \\
		$W^{+}W^{-}W^{\pm}$		     & $2.24\cdot 10^{-2}$        & $1.26\cdot 10^{-3}$       &    $\qquad 0.010 $&$^{+0.001}_{-0.001}$   \\
		$t\bar{t}+\mathrm{jets}$     & $8.67\cdot 10^{-4}$        & $1.89\cdot 10^{-4}$       &    $\qquad 3.6   $&$^{+0.2}_{-0.2}$     \\ 
		$t\bar{t}W^{\pm}j$		     & $2.44\cdot 10^{-2}$        & $1.11\cdot 10^{-2}$       &    $\qquad 0.37  $&$^{+0.01}_{-0.01}$     \\
		$t\bar{t}Z^{\pm}$		     & $1.67\cdot 10^{-2}$        & $8.24\cdot 10^{-3}$       &    $\qquad 0.17  $&$^{+0.01}_{-0.01}$     \\
		\hline                                                                                                                   
		BP~10	                     & $3.79\cdot 10^{-2}$	      & $2.87\cdot 10^{-2}$   	  &    $\qquad 61.8 $&$^{+1.1}_{-1.1}$     \\
		lBP~18                       & $4.97\cdot 10^{-2}$	      & $3.75\cdot 10^{-2}$       &    $\qquad 41.4 $&$^{+0.7}_{-0.7}$     \\							
		\end{tabular}
		\caption{The efficiency of the preselection cut, the total cut efficiency and number of expected 
		events for each background and the two signal samples for an integrated luminosity
		of $200 \ \mathrm{pb}^{-1}$. The stated uncertainty on the number of expected events
		corresponds to the 68.3\% confidence interval of this number. 
		The total background sums up to 6.7 events. 
		\label{table:events_after_cuts}}}
		
		\EPSFIGURE{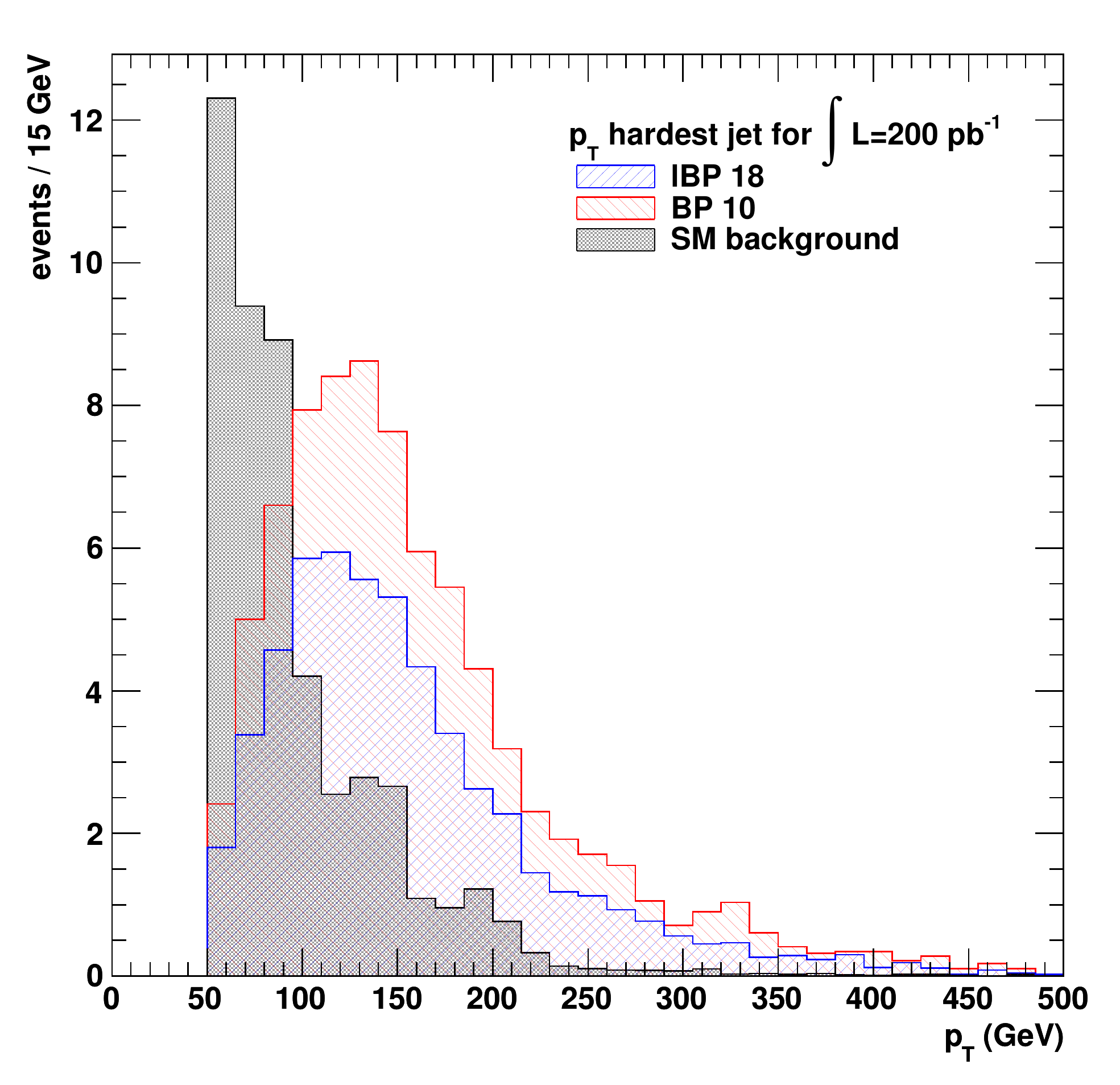,width=0.45\textwidth}
		{The $p_{T}$ of the hardest jet after preselection for BP~10 (red), 
		lBP~18 (blue) and the total SM background (grey).\label{fig:hardest_jet}}

		In table~\ref{table:events_after_cuts} we list the
		efficiencies of the preselection and of the superposition of all cuts for the two signal samples and for each 
		of the background processes. The efficiency of the cuts has been studied individually.		
		Moreover, we list the expected number of events after having superimposed 
		all cuts. 
		We find that we can expect 62 and 41 events for an integrated luminosity
		of $200 \ \mathrm{pb}^{-1}$ for BP~10 and lBP~18, but only a total of 6.7 events
		arising from the SM backgrounds. To estimate how much integrated luminosity we need to obtain
		a 5$\sigma$ excess over the SM expectation, we again make use of the log likelihood ratios.
		In particular, we calculate the background confidence level $1-\mathrm{CL_b}$ and require it to be smaller than 
		the 5$\sigma$ probability of $2.87 \cdot 10^{-7}$.  
		For lBP~18 we find a 5$\sigma$ significance for an integrated luminosity of $46^{+25}_{-22} \ \mathrm{pb}^{-1}$ 
		with an expected number of 
		$9.52^{+5.17}_{-4.55}$ signal and $1.54^{+0.84}_{-0.74}$ background events. For BP~10 we find that a 5$\sigma$ 
		excess is expected for an integrated luminosity 
		of $24^{+16}_{-12}$ $\mathrm{pb}^{-1}$, which corresponds to $7.42^{+4.94}_{-3.71}$ signal and $0.80^{+0.55}_{-0.40}$ background events. 
		The uncertainties are obtained in the same way as in section~\ref{sec:disc_pot}. 
		Systematic errors are not taken into account.
		These results show that a discovery of this model may already be feasible at the LHC with only a few dozen inverse 
		picobarns of understood collision data.


\section{Reconstructing the mass of a charge 5/3 top-partner}
\label{sec:mass_reco}

Among the top-partners, the charge 5/3 $x_1$ gives the largest contribution to the excess over the SM expectation 
in the SS di-lepton channel. This is due to its low mass and the fact that it always decays to $t W^{+}$, which
leads to 
\begin{equation}
	\label{eq:x1_decay}
	x_1 \to t W^{+} \to b W^{+}W^{+} \to b l^{+} l^{+} \nu_{l} \nu_{l}
\end{equation}
in the leptonic decay mode.
For $t_1$, which is the only new quark that could be lighter than $x_1$, 
only few of its decay modes (eq.~\eqref{eq:dominant_t1_x1_dec}) produce SS di-leptons in the final state. 
The accurate mass reconstruction of 
a charge 5/3 quark
would be a big step towards the interpretation 
of the discovery. In the literature, different methods have been proposed for the reconstruction of its mass.
These methods usually focus on pair production, so that they can exploit same-sign di-leptons 
from the decay of one of the charge 5/3 quarks to select and identify the event. The mass is reconstructed using 
the fully hadronic decay mode of the two $W$ bosons coming 
from the other charge 5/3 quark~\cite{AguilarSaavedra:2009es, CMS-PAS-EXO-08-008, Contino_08}.
In ref.~\cite{Mrazek:2009yu} an alternative method is presented. 
The mass of a charge 5/3 top-partner is reconstructed in SS di-lepton events
via its transverse mass. This transverse mass is computed from 
the momenta of the two SS leptons, the missing
transverse energy (from the two neutrinos) and the $b$ jet belonging to the semileptonically 
(and not to the second, hadronically) decaying top quark.

In the following, we outline a new method 
to reconstruct the mass of a charge 5/3 quark $x_1$. We exploit the same channel as~\cite{Mrazek:2009yu}, 
but we only rely on the two SS leptons and use the shape 
of their invariant mass distribution to reconstruct $x_1$. This 
avoids $b$ tagging inefficiencies and the problem of assigning the correct $b$ jet to the corresponding $x_1$ decay. 
We also consider  
the situation in which an excess of about 50 SS di-lepton events 
(as expected for 200~$\mathrm{pb}^{-1}$ of collision data) 
is caused by the presence of multiple top-partners. 
In this case, we show how the method can be used to discriminate the signal against a hypothesized presence of $x_1$ only.

\subsection{Mass determination with 200 pb$^{-1}$ of collision data}

	\paragraph{The method.} In the decay of a pair-produced $x_1\bar{x}_1$, the SS di-leptons come from the same decay leg and
	the positively (negatively) charged leptons can be assigned to the decay of $x_1$ ($\bar{x}_1$).
	The invariant mass distribution of the SS di-leptons contains information about the $x_1$ mass. 
	In fact, the endpoint of this invariant mass distribution $m_{ll}^{\mathrm{max}}$ is sufficient 
	to determine $m_{x_1}$, since $m_{ll}^{\mathrm{max}}$ can be expressed in terms of the masses of the particles
	involved in the decay \eqref{eq:x1_decay}.
	The mass of $x_1$ is the only unknown parameter in this relation. 
	An accurate measurement of this endpoint, however, is not possible with only 200~$\mathrm{pb}^{-1}$ of collision data. 
	We can use, instead, the shape of the invariant mass distribution to determine $m_{x_1}$.  	
		
	In ref.~\cite{Raklev}, an analytic expression for the shape of the invariant mass distribution $M_{lc}$ for the 
	supersymmetric decay $\tilde{g}\to \bar{t} \ \tilde{t}_{1}$, $\tilde{t}_{1} \to c \ \tilde{\chi}_{1}^{0}$ 
	is presented\footnote{Spin effects were neglected in the calculation of 
	the shape of the invariant mass 
	distribution.}. 
	As the kinematic configuration of this decay is identical to eq.~\eqref{eq:x1_decay}, 
	we can use their results to model the shape of the invariant mass distribution of 
	the SS di-leptons from leptonic $x_1$ ($\bar{x}_1$) decays. 
	This shape function, however, does not take into account the
	possibility of a leptonically decaying tau-lepton originating from a $W$ decay.	
	Also, an inclusive electron and muon	
	spectrum 
	without any selection cuts was assumed. These two assumptions are not satisfied in our realistic analysis. 
	A fit of the full invariant mass distribution does therefore not
	lead to an accurate estimation of $m_{x_1}$. However, we find the shape of the tail of the distribution
	to be almost invariant under the effect of 
	the selection cuts and the tau contribution\footnote{The systematic error introduced is $< 3\%$.}. 
	A fit of the tail of the distribution is thus a powerful means to extract the mass of $x_1$. 
	
	In figure \ref{fig:x1_fit} 
	we show the invariant mass distribution of the SS di-leptons
	from a pair-produced $x_1$ with a mass of 365~$\mathrm{GeV}$.
	This is the $x_1$ mass for BP~10 and lBP~18. We apply the same selection 
	as in section~\ref{sec:two_smoking_guns}. 
	Both the signal and the SM background (see table~\ref{table:events_after_cuts}) 
	are normalized to an integrated luminosity of 200~$\mathrm{pb}^{-1}$.
	With a leading order cross section of 1.64~$\mathrm{pb}$ for the signal, we estimate $15.3 \pm 0.2$ SS di-lepton events
	due to $x_1\bar{x}_1$.
	When fitting the tail of the total distribution from the signal plus the SM 
	with the shape function (starting from the peak of the distribution), we 
	obtain a fitted mass $m_{\mathrm{fit}}$ of $370.0 \pm 32.3 \ \mathrm{GeV}$. By rescaling the generated distribution
	with the signal cross section, we underestimate the statistical fluctuations in the number of events per bin. 
	The statistical uncertainty of about 32~GeV on the fitted mass, however, correctly represents the precision expected with 
	about 15 signal events. 
	We conclude that fitting the tail of the invariant mass distribution of the signal plus the SM background 
	leads to a fairly accurate estimate of the $x_1$ 
	mass\footnote{When only fitting the signal without the SM background, we find $m_{\mathrm{fit}}=372.4 \pm 30.3$~GeV.}.
	
	\EPSFIGURE{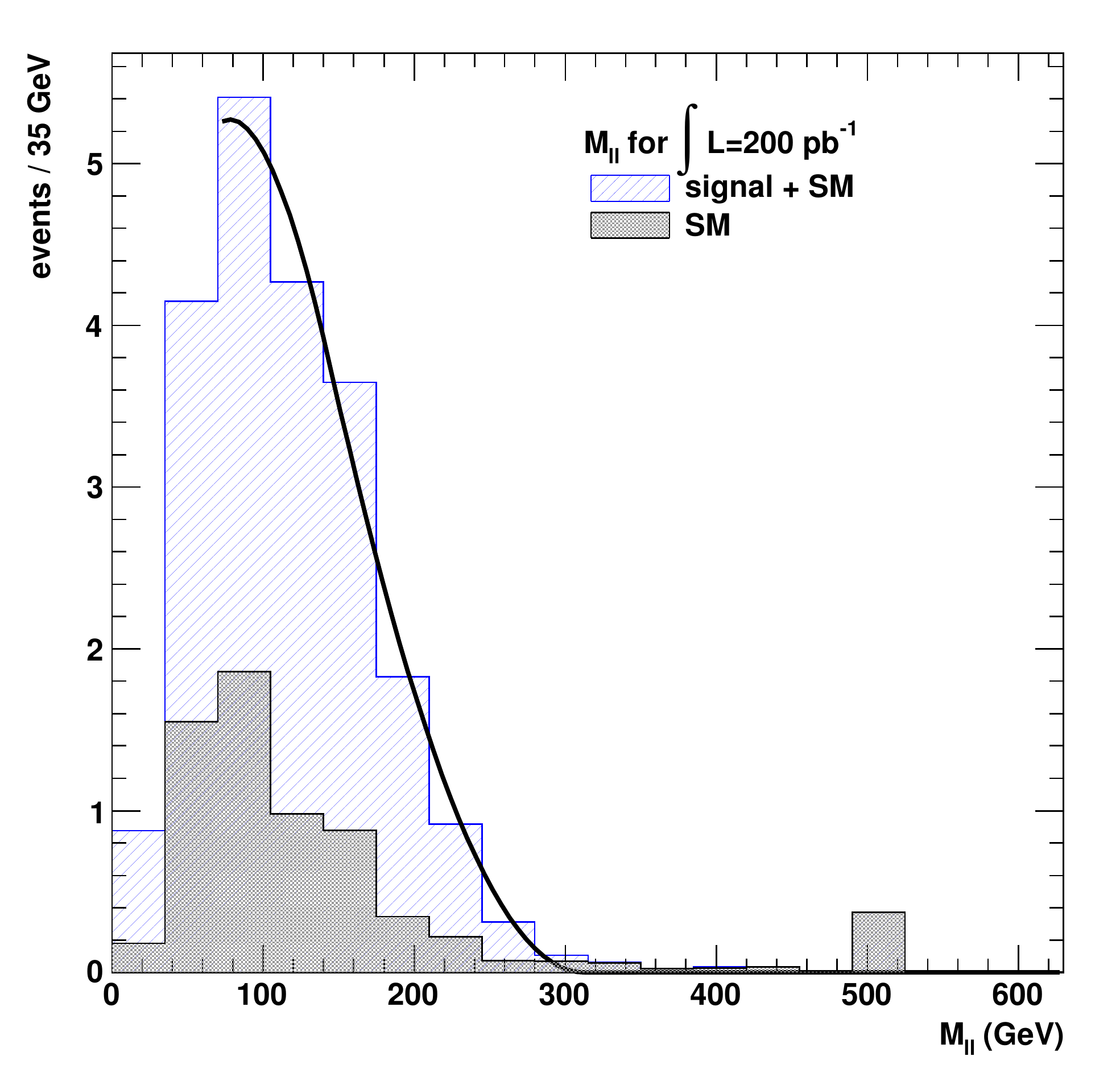,width=0.45\textwidth}
	{Invariant mass of the SS di-leptons from leptonic decays of a $x_1$ with mass 365~$\mathrm{GeV}$.
	The signal is stacked on top of the SM. 
	The peak in the SM distribution at an invariant mass 
	of about 520~GeV is caused by a single $Z\to l^{+}l^{-}$ event that passed the selection cuts.\label{fig:x1_fit}}	
	
	The above method assumes the total production cross section of the charge 5/3 top-partners to be dominated by pair-production. 
	Neglecting the contribution of single quark production allows us to estimate the production cross section as a function
	of the quark mass. This neglected contribution affects neither the shape nor the endpoint, but
	changes the absolute normalization of the 
	invariant mass distribution of the SS di-leptons.   
	The cross section for single quark production is typically small for relatively light top-partners, 
	but influenced by model-dependent electroweak couplings. For BP~10 and lBP~18 we find that the ratio of the 
	leading order cross section for single $x_1$ production over $x_1\bar{x}_1$ pair production is about 5.8\% and 2.3\%, respectively.
	For these points, the errors introduced are smaller than the uncertainty of the next-to-leading order pair production cross section,
	which is approximately 20\% for top-partners with a mass of about 500~GeV \cite{Berger:2009qy}.
	
	\paragraph{Applying the method to two benchmark points.}
	
	For the $\mathbf{4}$ of $SO(4)$ and the $XX$ signatures, the various top-partners in addition to $x_1$ contribute to the 
	excess of SS di-lepton events and alter the invariant mass distribution. 
	In the special case of BP~10 and lBP~18, 
	there is a bottom-like $b_1$ with a mass of about 10~GeV above the $x_1$ mass. 
	Since it predominantly decays to $W^{-}t$, it plays an important role for the additional production of SS di-leptons. 
	In order to obtain SS (rather than OS) di-leptons from $b_1\bar{b}_1$ decays, 
	one lepton has to come from $b_1$ and the other from $\bar{b}_1$. 
	Therefore, the invariant mass distribution of the SS di-leptons from $b_1\bar{b}_1$ decays does not show an endpoint, 
	but rather a tail that extends far into the high invariant mass region. This is in contrast to the SS di-leptons from $x_1\bar{x}_1$ decays. 
	In case of BP~10, two charge
	5/3 quarks below 500~GeV contribute to the excess of SS di-lepton events. Since $x_2$ is 
	more massive than $x_1$, the invariant mass distribution due to its leptonic decay 
	is broader and has a larger endpoint with respect to the $x_1$ contribution. 
	The main effects of these additional top-partners (including the charge 2/3 quarks) 
	on the invariant mass distribution of the SS di-leptons are an 
	increased number of signal events and a large tail that hides the endpoint due to the light $x_1$. These effects can be used
	to determine whether or not the expected SS di-lepton invariant mass distributions for BP~10 and lBP~18 can 
	be explained by the hypothesized presence of a charge 5/3 top-partner only. 
	
	In figures~\ref{fig:BP10_fitted} and~\ref{fig:lBP18_fitted}
	we show the invariant mass distribution of the SS di-leptons for BP~10 and lBP~18 respectively,
	as expected to be observed with 200~$\mathrm{pb}^{-1}$ of collision data. 
	The SM background for the same integrated luminosity is added to the signal distribution. As explained above, 
	a fit of the tail of the distribution leads to a fairly accurate estimate of the $x_1$ mass, if
	the observation is caused by only one charge 5/3 quark (plus the SM contribution). 
	For BP~10, we obtain $m_{\mathrm{fit}}=395.5\pm24.6$~GeV and for lBP~18, we find $m_{\mathrm{fit}}=388.6\pm29.7$~GeV. 
	This shows that a fit of the total distribution, 
	including the contributions from the various top-partners and the SM backgrounds, 
	leads to a systematic overestimate of the mass, 
	which nevertheless remains within about 1$\sigma$ of the true $x_1$ mass. 
	
	\DOUBLEFIGURE{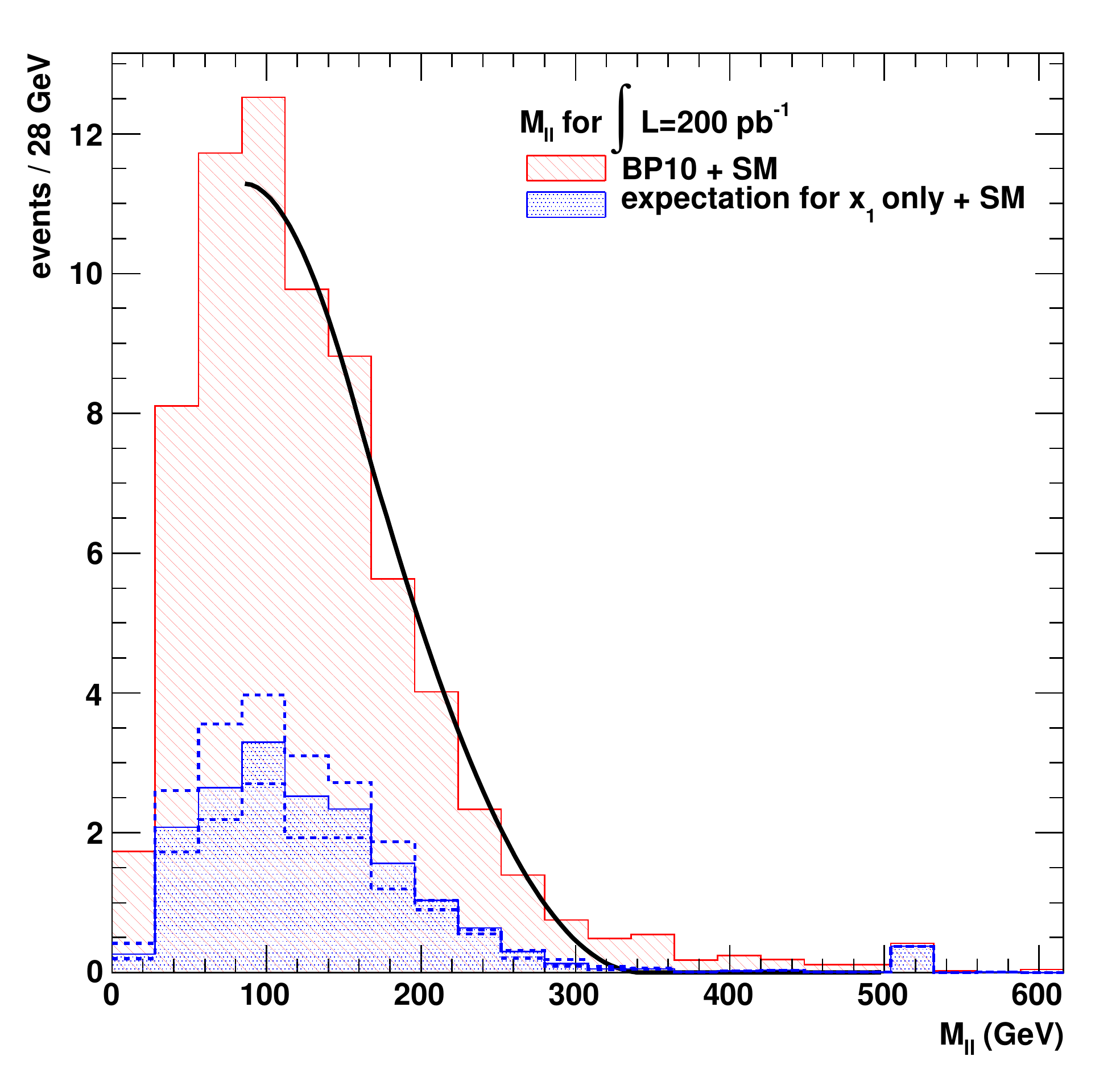,width=0.45\textwidth}
	{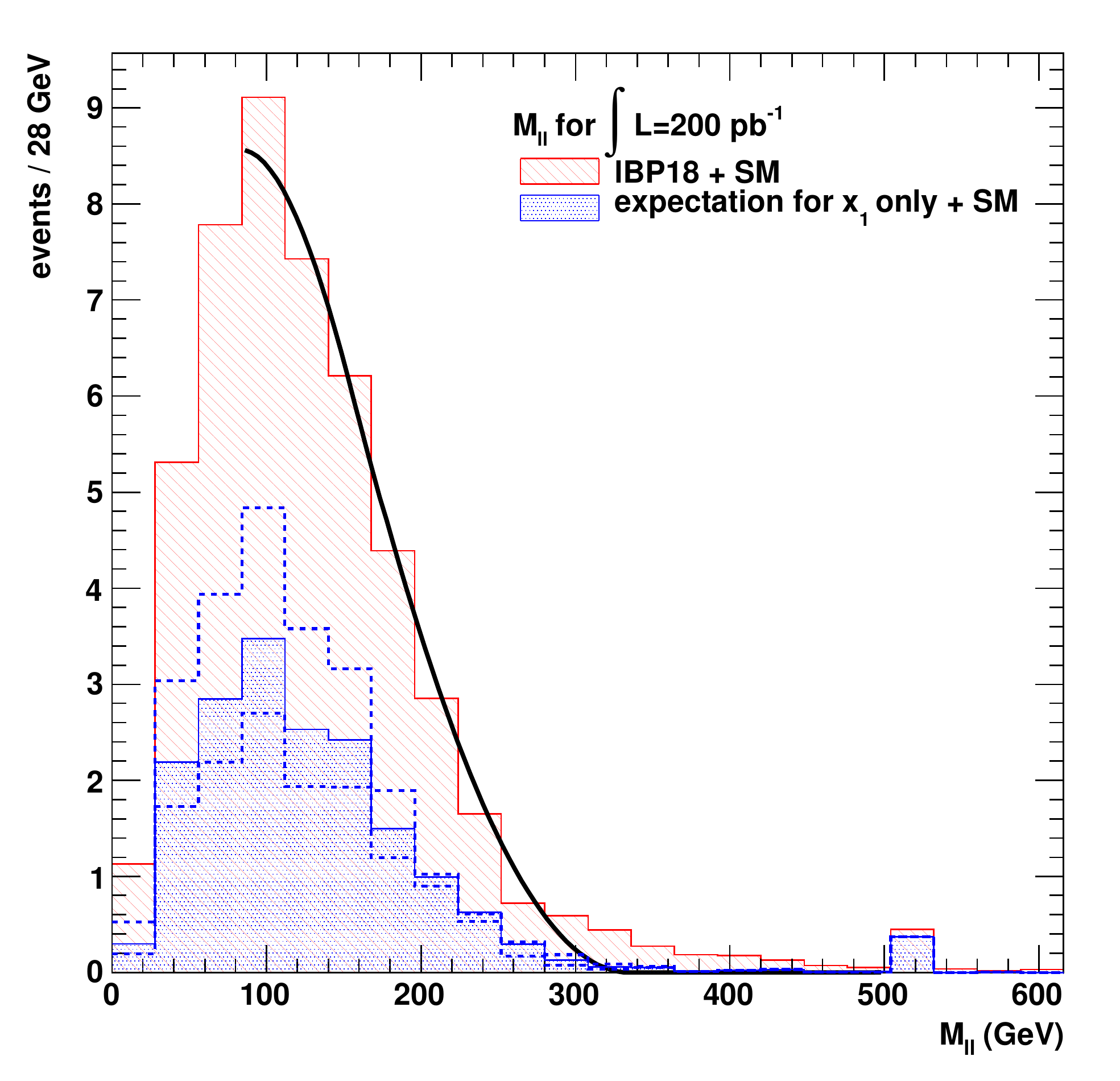,width=0.45\textwidth}
	{The invariant mass distribution of the SS di-leptons for BP~10 plus the SM background (red), overlaid with
	the expected distribution due to a pair-produced $x_1$ quark with a mass $m_{\mathrm{fit}}=395.5$~GeV (blue). 
	The (blue) dashed distributions correspond to the variation of the $x_1$ expectation 
	due to the uncertainty on the fitted mass of 24.6~GeV. 
	\label{fig:BP10_fitted}}
	{Same distributions as on the left, but for lBP~18 with $m_{\mathrm{fit}}=388.6\pm29.7$~GeV. \label{fig:lBP18_fitted}}
	
	As a next step, we calculate the cross section and simulate the expected
	signal for a pair-produced $x_1$ with the fitted masses. 
	This signal plus the SM expectation gives the expected invariant mass distribution
	of the SS di-leptons for a given mass hypothesis.
	For BP~10 and lBP~18 (figures~\ref{fig:BP10_fitted} and~\ref{fig:lBP18_fitted}), we see that neither 
	of the two signal distributions can be explained assuming the presence of only one charge 5/3 quark. 
	In particular, we expect 62.5 and 6.7 SS di-leptons from BP~10 and the SM background respectively.
	Fitting the tail of the SS di-lepton invariant 
	mass distribution, however, leads to an estimate of $10.6^{+3.3}_{-2.9}$ SS di-leptons due to a pair-produced 
	charge 5/3 top-partner. The stated errors are due to the statistical 
	uncertainty on the fitted mass only. For lBP~18, $42.4$ SS di-leptons are expected 
	from the top-partners with 200~$\mathrm{pb}^{-1}$. Assuming the presence of $x_1$ only, the tail of the distribution
	suggests an $x_1$ mass that can account for $11.2^{+5.4}_{-3.4}$ SS di-leptons. For BP~10
	and lBP~18 we are left with respectively 52 and 31 unexplained SS di-leptons. The uncertainty on these numbers 
	is dominated by the Poisson uncertainty of the 62.5 and 42.4 expected events.

	The possibility of the signal to be mainly caused by a very light $t_1$ can be excluded in the following way. 
	The dominant channel for $t_1$ to produce SS di-lepton events includes 
	the leptonic decay of a $Z$, $t_1 \to t Z$. In this case, three leptons are produced and two OS leptons come from the $Z$. 
	A veto on a mass window around $m_Z$ for OS di-leptons thus helps to suppress the $t_1$ contribution to the signal. 
	For BP~10 and lBP~18, leptonic $t_1$ decays account for 5\% and 19\% of the signal. 
	Cutting on a window of $m_Z\pm10$~GeV results in a loss of about 20\% of the total signal, but reduces
	the $t_1$ contribution by about 70\%. Alternatively, one could directly veto tri-lepton final 
	states to curb the contribution from $t_1$.
	
	We conclude that for both benchmark points, 200~$\mathrm{pb}^{-1}$ of collision data would be sufficient to obtain an evident
	discrepancy between the total invariant mass distribution and the distribution based
	on the hypothesized presence of only one charge 5/3 top-partner. Such an observation could be seen as evidence in favour of 
	a model with multiple top-partners. 
	If instead the signal distribution were consistent with the 
	expected distribution from $m_\mathrm{fit}$,
	much more than 200~$\mathrm{pb}^{-1}$ of collision data would be
	needed for the distribution to reveal the presence of additional, heavier top-partners.
	
	\paragraph{Beyond the 200~$\mathbf{pb}^{-1}$ scenario.}
	When more integrated luminosity has been collected at the LHC, advanced techniques 
	can be used to resolve more details about the masses of the top-partners.
	The identification of either a full \textbf{4} of $SO(4)$ or two charge 5/3 
	quarks would be evidence in favour of our model. 
	The signal in the SS di-lepton channel can be produced by various top-partners and it may be 
	difficult to disentangle the different contributions. Discriminating 
	the SS di-lepton events due to $x_1$ from the contribution due to $b_1\bar{b}_1$ would be an important step. 
	In the SS di-lepton channel, the two leptons from $x_1\bar{x}_1$ decays come from the same particle, 
	whereas in $b_1\bar{b}_1$ decays they come one from a quark each. In the OS di-lepton channel, the roles 
	of $x_1$ and $b_1$ are exchanged. The shapes of the SS and OS di-lepton invariant mass distributions may help
	to gain insight in the underlying physics.

	
\section{Conclusions}
\label{conclusions}

We reviewed a composite Higgs model and 
highlighted some of its most important features. 
We used vector-like fermionic resonances to 
reconcile the model with EWPT. This
is more easily achieved when two sets of composites 
are below the cutoff of the effective theory.
We showed that in this case the collider 
phenomenology is very rich, and in particular
we can obtain some distinctive signatures for our model. 
These are the cases when a full $\mathbf{4}$ of $SO(4)$ 
or two charge 5/3 top-partners lie within the reach of the LHC.
We scanned the parameter space of the model focussing on points 
that are consistent with EWPT observables and give these signatures.
For these signatures we described the possible mass hierarchies and 
outlined the basic features of their phenomenology.
We find that the tri-lepton and same-sign di-lepton final states are 
the most promising ones for a discovery of the model.

We studied in detail the phenomenology of two 
benchmark points with a large production cross section.
Both exhibit a $\mathbf{4}$ signature and one has 
two charge 5/3 quarks with a mass below 500 GeV. 
We presented a robust cut-based search strategy for an excess in final states
with at least two same-sign leptons. After making a basic kinematic selection, 
only little background from the SM was found in this channel. 
We find that for both benchmark points a few tens of inverse
picobarns of understood collision data would suffice to observe
a 5$\sigma$ significance.

Since the SM contamination in the same-sign di-lepton final state
is small, this channel is not only well 
suited for observing an excess over the SM expectation, 
but also for reconstructing the masses of 
the new particles. Among the top-partners, 
the light charge 5/3 quark contributes the most 
to the excess of SS di-lepton events. 
We described a new method to reconstruct 
the mass of such a quark via its leptonic decay. 
This method only relies on the reconstruction
of the two same-sign leptons and exploits 
the shape of their invariant mass distribution.
For both distinctive signatures of the model, the light top-partners 
besides the charge 5/3 quark also contribute
to the excess of same-sign di-lepton events. 
In this case, we showed how the mass reconstruction method could be used to judge if the
excess of same-sign di-lepton events is compatible with 
the presence of a charge 5/3 quark only, or if it hints at the existence 
of additional top-partners.
For this, we used the fact that the cross section for 
pair production of top-partners can be predicted as a function of mass. 
Already with an integrated luminosity of 
200~$\mathrm{pb}^{-1}$ and a corresponding statistics of about 50 signal events, 
we found an evident disagreement 
between the single $x_1$ hypothesis and the expected observation. Such a disagreement 
can be seen as an indication for the presence 
of top-partners in addition to a charge 5/3 quark.


\section*{Acknowledgments}
We are indebted to Charalampos Anastasiou, Luc Pape and Daniel Treille for their advice and encouragement. 
We also wish to thank Jose Santiago and Fabian Stoeckli for inspiring discussions and Giuliano Panico
for his comments on the draft. We are grateful to Rikkert Frederix for his assistance on the use of MadGraph.
This research is supported by the Swiss National
Science Foundation under contracts 200020-126632 and PDFMP2\_127462. 

\appendix
\section{Discovery potential of the 30 benchmark points} 
\label{sec:CLb_table_appendix}

In table \ref{table:CLb}, 
we list the number of expected events for the 30 benchmark points in each of the four considered final states. 
We also give the corresponding background confidence level~$1-\mathrm{CL_b}$ arising from 
a combined search in the four channels. We use the log likelihood ratios (as defined in section \ref{sec:disc_pot}) as a test statistic.
We indicate if the central $1-\mathrm{CL_{b}}$ value corresponds to an excess of at least 3$\sigma$ or 5$\sigma$.
\TABLE{\begin{tabular}{@{}rccccr@{\hspace{3mm}}lr@{}}
BP				&	2l SS + 3j  &  2l SS + 4j & 3l OS + 2j & 3l OS + 3j	&	\multicolumn{3}{c}{$\mathrm{1-CL_{b}}$}                                      \\   
\hline                                                                                                                                                      
lBP 1			&	7.22   	&	5.82  		&4.45	   &4.12      &	$ 1.3         \cdot10^{-6} $&$^{+2.2  \cdot10^{-4} }_{- 1.3  \cdot10^{-6}}$     &  $>3\sigma$\\    
hBP 1   		&	3.76  	&	3.40	  	&2.35      &2.53      &	$ 2.2         \cdot10^{-3} $&$^{+4.1  \cdot10^{-2} }_{- 2.2  \cdot10^{-3}}$     &  \\              
lBP 2			&	6.70  	&	5.59	  	&4.13	   &3.42      &	$ 5.9         \cdot10^{-6} $&$^{+6.8  \cdot10^{-4} }_{- 5.9  \cdot10^{-6}}$     &  $>3\sigma$\\    
hBP 2   		&	4.64  	&	4.24     	&2.68	   &2.71      &	$ 4.8         \cdot10^{-4} $&$^{+1.5  \cdot10^{-2} }_{- 4.7  \cdot10^{-4}}$     &  $>3\sigma$\\    
lBP 3			&  12.70 	&	9.67	   	&8.49	   &7.27      &	$ ^{\ast}\ 1  \cdot10^{-9} $&$^{                   }_{                  }$     &  $>5\sigma$\\    
hBP 3   		&	6.61  	&	5.87   	    &4.22	   &4.18      &	$ 2.4         \cdot10^{-6} $&$^{+3.4  \cdot10^{-4} }_{- 2.4  \cdot10^{-6}}$     &  $>3\sigma$\\    
lBP 4			&	8.10  	&	7.01   	    &4.65	   &4.38      &	$ 7.9         \cdot10^{-8} $&$^{+3.2  \cdot10^{-5} }_{- 7.9  \cdot10^{-8}}$     &  $>5\sigma$\\    
hBP 4 			&	2.94  	&	2.85   	    &1.70	   &1.78      &	$ 1.1         \cdot10^{-2} $&$^{+1.1  \cdot10^{-1} }_{- 1.1  \cdot10^{-2}}$     &            \\    
BP 5			&	6.97  	&	5.65   	    &3.93	   &3.80      &	$ 3.7         \cdot10^{-6} $&$^{+4.9  \cdot10^{-4} }_{- 3.7  \cdot10^{-6}}$     &  $>3\sigma$\\	  
BP 6			&	2.91  	&	2.59   	    &1.73	   &1.78      &	$ 1.4         \cdot10^{-2} $&$^{+1.2  \cdot10^{-1} }_{- 1.3  \cdot10^{-2}}$     &  \\	          
BP 7			&	3.06  	&	3.03   	    &1.73	   &1.81      &	$ 9.1         \cdot10^{-3} $&$^{+1.0  \cdot10^{-1} }_{- 8.8  \cdot10^{-3}}$     &  \\	          
BP 8			&	2.98  	&	2.81   	    &1.50	   &1.52      &	$ 1.4         \cdot10^{-2} $&$^{+1.3  \cdot10^{-1} }_{- 1.3  \cdot10^{-2}}$     &  \\	          
BP 9			&	7.19  	&	5.66   	    &3.81	   &3.68      &	$ 3.8         \cdot10^{-6} $&$^{+4.9  \cdot10^{-4} }_{- 3.8  \cdot10^{-6}}$     &  $>3\sigma$\\	  
BP 10			&  14.62 	&	10.87      	&9.14      &7.59      &	$ ^{\ast}\ 1  \cdot10^{-9} $&$^{                   }_{                   }$     &  $>5\sigma$\\	  
lBP 11			&  10.37 	&	8.63   	    &5.56	   &5.40      &	$ ^{\ast}\ 1  \cdot10^{-9} $&$^{+2.7  \cdot10^{-7} }_{                   }$     &  $>5\sigma$\\	  
hBP 11        	&	5.70  	&	5.08   	    &3.32	   &3.63      &	$ 3.4         \cdot10^{-5} $&$^{+2.4  \cdot10^{-3} }_{- 3.4  \cdot10^{-5}}$     &  $>3\sigma$\\	  
lBP 12			&	7.87  	&	5.87   	    &4.20	   &3.53      &	$ 1.5         \cdot10^{-6} $&$^{+2.3  \cdot10^{-4} }_{- 1.5  \cdot10^{-6}}$     &  $>3\sigma$\\	  
hBP 12        	&	8.42  	&	7.30   	    &4.86	   &4.72      &	$ 3.7         \cdot10^{-8} $&$^{+1.3  \cdot10^{-5} }_{- 3.7  \cdot10^{-8}}$     &  $>5\sigma$\\	  
BP 13			&  12.98 	&	10.16      	&8.74	   &7.38      &	$ ^{\ast}\ 1  \cdot10^{-9} $&$^{                   }_{                   }$     &  $>5\sigma$\\	  
BP 14			&  10.44 	&	8.83   	    &6.59	   &6.74      &	$ ^{\ast}\ 1  \cdot10^{-9} $&$^{+2.2  \cdot10^{-8} }_{                   }$     &  $>5\sigma$\\	  
lBP 15			&	7.99  	&	6.11   	    &4.12	   &3.83      &	$ 8.4         \cdot10^{-7} $&$^{+1.5  \cdot10^{-4} }_{- 8.4  \cdot10^{-7}}$     &  $>3\sigma$\\	  
hBP 15        	&	6.89  	&	5.71   	    &4.24	   &3.99      &	$ 2.5         \cdot10^{-6} $&$^{+3.6  \cdot10^{-4} }_{- 2.5  \cdot10^{-6}}$     &  $>3\sigma$\\	  
lBP 16			&	8.57  	&	6.87   	    &5.74	   &4.86      &	$ 2.3         \cdot10^{-8} $&$^{+8.6  \cdot10^{-6} }_{- 2.3  \cdot10^{-8}}$     &  $>5\sigma$\\	  
hBP 16        	&	2.32  	&	2.21   	    &1.42	   &1.34      &	$ 3.2         \cdot10^{-2} $&$^{+2.0  \cdot10^{-1} }_{- 3.0  \cdot10^{-2}}$     &           \\	  
lBP 17			&	9.15  	&	6.72   	    &5.44	   &4.83      &	$ 2.0         \cdot10^{-8} $&$^{+8.2  \cdot10^{-6} }_{- 2.0  \cdot10^{-8}}$     &  $>5\sigma$\\	  
hBP 17        	&	2.43  	&	2.17   	    &1.35	   &1.40      &	$ 3.2         \cdot10^{-2} $&$^{+2.0  \cdot10^{-1} }_{- 3.0  \cdot10^{-2}}$     &  \\	          
lBP 18			&  10.03 	&	7.48   	    &6.53	   &5.09      &	$ ^{\ast}\ 1  \cdot10^{-9} $&$^{+6.8  \cdot10^{-7} }_{                   }$     &  $>5\sigma$\\	  
hBP 18        	&	4.42  	&	3.77   	    &2.67	   &2.44      &	$ 9.7         \cdot10^{-4} $&$^{+2.5  \cdot10^{-2} }_{- 9.6  \cdot10^{-4}}$     &  $>3\sigma$\\	  
BP 19			&	7.31  	&	5.42   	    &4.09	   &3.83      &	$ 3.2         \cdot10^{-6} $&$^{+4.2  \cdot10^{-4} }_{- 3.1  \cdot10^{-6}}$     &  $>3\sigma$\\	  
BP 20       	&  	6.49 	&   5.06   	    &3.79	   &3.75      &	$ 1.3         \cdot10^{-5} $&$^{+1.1  \cdot10^{-3} }_{- 1.3  \cdot10^{-5}}$     &  $>3\sigma$\\	   			
\end{tabular}
\caption{Number of expected events in  each of the four channels for
200~pb$^{-1}$ of integrated luminosity. The background confidence level 
$1-\mathrm{CL_{b}}$ with its uncertainty is also given. 
The $1-\mathrm{CL_b}$ values marked with $(^\ast)$ correspond to benchmark points for which more than $10^{9}$ pseudoexperiments would be
needed for the tail of the tPDF of the background hypothesis to leak out of the integrated region.
\label{table:CLb}}
}
\noindent

\newpage

\end{document}